\documentclass[lettersize,journal]{IEEEtran}
\usepackage{amsmath,amsfonts}
\usepackage{algorithmic}
\usepackage{algorithm}
\usepackage{array}
\usepackage[caption=false,font=normalsize,labelfont=sf,textfont=sf]{subfig}
\usepackage{textcomp}
\usepackage{stfloats}
\usepackage{url}
\usepackage{verbatim}
\usepackage{graphicx}
\usepackage{cite}
\usepackage{makecell}
\usepackage{multirow}
\begin{document}

\title{Neural Networks Enabled Discovery On the Higher-Order Nonlinear Partial Differential Equation of Traffic Dynamics}

\author{Zihang Wei, Yunlong Zhang, Chenxi Liu, Yang Zhou, \IEEEmembership{Member, IEEE},
\thanks{Zihang Wei, Yunlong Zhang and Yang Zhou are with the Zachry Department of Civil \& Environmental Engineering, Texas A\&M University, College Station, Texas, USA (e-mail: wzh96@tamu.edu; yangzhou295@tamu.edu; yzhang@civil.tamu.edu) Chenxi Liu is with the Department of Civil \& Environmental Engineering, University of Utah, Utah, USA. (e-mail: chenxi.liu@utah.edu) (Corresponding Author: Yang Zhou)}
} 

\markboth{IEEE Transitions on Intelligent Transportation Systems}%
{Shell \MakeLowercase{\textit{et al.}}: A Sample Article Using IEEEtran.cls for IEEE Journals}


\maketitle

\begin{abstract}
Modeling the traffic dynamics is essential for understanding and predicting the traffic spatiotemporal evolution. However, deriving the partial differential equation (PDE) models that capture these dynamics is challenging due to their potential high order property and nonlinearity. In this paper, we introduce a novel deep learning framework—TRAFFIC-PDE-LEARN—designed to discover hidden PDE models of traffic network dynamics directly from measurement data. By harnessing the power of the neural network to approximate a spatiotemporal fundamental diagram that facilitates smooth estimation of partial derivatives with low-resolution loop detector data. Furthermore, the use of automatic differentiation enables efficient computation of the necessary partial derivatives through the chain and product rules, while sparse regression techniques facilitate the precise identification of physically interpretable PDE components. Tested on data from a real-world traffic network, our model demonstrates that the underlying PDEs governing traffic dynamics are both high-order and nonlinear. By leveraging the learned dynamics for prediction purposes, the results underscore the effectiveness of our approach and its potential to advance intelligent transportation systems.
\end{abstract}


\section{Introduction}
\label{intro}
\IEEEPARstart{T}{he} modeling of traffic network dynamics is essential for understanding the spatiotemporal evolution laws of the traffic network. This is indispensable to the modern application of intelligent transportation system (ITS) including accurately predicting and controlling the traffic states. Studies on macroscopic traffic flow modeling commonly utilize partial differential equations (PDEs). This type of equations typically models the temporal evolutions of traffic measurements as a function of the spatial evolutions of traffic measurements. Conventional approaches to deriving macroscopic traffic PDE models typically rely on a foundation of well-established physical and mathematical laws. However, in real-world application, due to the inherent complexity of driving behaviors and traffic network geometries, traffic PDE models tend to demonstrate higher-order and nonlinear property \cite{vlahogianni_temporal_2008, papageorgiou_dynamic_1990, aw2000resurrection, zhang2002non}.  This will inevitably increase the challenges of accurately discover reliable macroscopic traffic models via PDEs in a network. 


One of the earliest macroscopic traffic flow model can be found in the 1950s following the introduction of the Lighthill-Whitham-Richards (LWR) model \cite{lighthill1955kinematic2, richards1956shock}. The LWR model is developed based on the flow conservation law and represents the traffic kinematic wave theory as $\partial_tk + \partial_xq = 0$, where $k$ and $q$ denote the traffic density and flow respectively. Furthermore, the LWR model assumes that the traffic state is equilibrated following a fundamental diagram (FD) \cite{greenshields1935study, smulders_control_1990, drake_statistical_1966, newell_nonlinear_1961}  according to which traffic flow $q$ can be written as a function of density $k$ as $q=Q(k)$. The LWR PDE can then be reformulated as $\partial_tk + \partial_xQ(k) = 0$ which describes the temporal evolution of traffic density $k$ on a single highway of infinite length. Later, Daganzo \cite{daganzo1994cell} has demonstrated that the LWR model can be reformulated as a discretized version namely the cell transmission model (CTM). The CTM considers the traffic network traffic network as multiple connected discretized cells. The density of a certain cell is updated to the next timestep as its current timestep’s density plus the inflow to this cell and minus the outflow to the next cell. It has been proved that, with proper choice of the timestep length, cell distance, and the flow-density FD, the CTM is proved to be equivalent to a discretized version of the LWR model. The LWR and CTM are among the most popular macroscopic traffic models due to its simplicity and their ability to model simple traffic networks. 


Furthermore, several variants of CTM have been proposed, including stochastic CTM (SCTM) \cite{sumalee_stochastic_2011}, asymmetric CTM (ACTM) \cite{gomes_optimal_2006}, and CTM with a linearly decreasing function \cite{srivastava_modified_2015}, to model traffic dynamics under various complex traffic conditions. Nevertheless, the choice of spatiotemporal step sizes for the macroscopic traffic network PDE models are required to be explicitly defined to ensure the models’ legitimacy. For instance, in the LWR model, the spatial and temporal step sizes should satisfy $\partial_x=c\partial_t$ where $c$ is the wave velocity \cite{lighthill1955kinematic1}, and the CTM model requires users to determine the proper cell length and clock tick \cite{daganzo1994cell}. This can limit the effectiveness of these macroscopic traffic network PDE models as the wave velocity can change along a highway and the traffic measurement data collected from real-world traffic network may not have the desired space step and time step.

Another limitation is related to the assumption that traffic always follows the equilibrium state defined by the FDs. Under which traffic flow reacts instantaneously to density changes. This assumption essentially neglects the existence of stop-and-go traffic caused by congestion under which speed and density do not follow FDs in equilibrium states \cite{yu2019traffic}. To capture more accurate dynamics,  second-order models, which contain two differential equations, have been proposed in order to tackle this limitation, including the Payne-Whitham (PW) model \cite{payne1971model, whitham2011linear} and the Aw-Rascle-Zhang (ARZ) model \cite{aw2000resurrection, zhang2002non}. The PW model developed a speed equation in order to tolerate deviations from the speed-density equilibrium. However, according to the critics brought up by Daganzo \cite{daganzo1995requiem}, the PW model is based on fluid flow theory which considers vehicles as isotropic fluid particles that can be affected by vehicles from both behind and the front. While vehicles are anisotropic particles that only respond to vehicles in the front. Later, Aw and Rascle \cite{aw2000resurrection} and Zhang \cite{zhang2002non} separately proposed new velocity equations to address this issue resulting in the ARZ model. However, the introduced new velocity equation may not fully capture the nuances of driving behavior,  \cite{yu2019traffic,fan_heterogeneous_2015,corli_hysteretic_2024}. There are other factors, complex segment geometries, that can further hinder the accuracy of existing PDE models \cite{buisson2009exploring, seo2017traffic}.


Although efforts have been made to address the abovementioned limitations in order to improve the macroscopic traffic network PDE models, these models are still inevitably relying on prior knowledge and pre-assumed driving behaviors. However, due to the complexity of most real-word traffic networks, their prior knowledge is challenging to obtain and often limited which will lead to inaccurate first principles. For example, in order to implement the LWR model with on-ramp/off-ramp source terms, exact knowledge regarding the traffic inflow and outflow through ramps are necessary while may not always be available in practice. Moreover, even though prior knowledge of complex traffic networks is available, the hidden PDEs of complex macroscopic traffic networks tend to demonstrate higher-order and nonlinear properties. This indicates that the PDE models can become extremely complicated, and we may not be able to rigorously derive them in an easy way.

Given the limitations and challenges of the existing macroscopic traffic network PDE models, the study aims to propose a data-driven method in order to discover the higher-order nonlinear hidden PDEs of complex macroscopic traffic network dynamics. Previous studies have proposed physics-informed models to uncover the underlying traffic network dynamics \cite{wang_pi-stgnet_2024, wang_koopman_2024}. Furthermore, studies have tried to discover the ordinary differential equations (ODEs) of traffic network dynamics based on observed data \cite{wei_discover_2025, wang_anti-circulant_2023} and applied them for network control tasks \cite{wei_coordinated_2025, gu_deep_2023}. In recent years, with the increasing availability of data collected from various dynamical systems, the data-driven discovery of hidden PDEs is experiencing a rapid development. The first major breakthrough is the introduction of the “PDE-FIND” model \cite{rudy2017data}, this study utilized data collected from dynamical systems on a spatiotemporal grid with fixed interval and discover the hidden PDEs using sparse regression technique based on a collection of potential function terms. In a later study, Schaeffer \cite{schaeffer2017learning} proposed a similar method to discover PDE from data. However, since data directly collected from the systems contain noisy measurements and moreover, it has been found that the derivatives calculated from the noisy measurements can further amplify the noise, studies started to apply deep neural networks to take coordinates as inputs and approximate the noisy measurements as outputs \cite{berg2019data, both2021deepmod, xu2021deep}. This improvement has been proved to effectively denoise the measured data and the derivatives can be directly computed through automatic differentiation technique \cite{baydin2018automatic,stephany2022pde, stephany2024pde, stephany_weak-pde-learn_2024}. 

To this end, existing traffic network PDEs models have limitations in terms of their strong reliance on prior knowledge and replication of human driving behaviors, which hinders their applicability on complex real-world traffic networks. To bridge this gap, this study proposes a novel model named “TRAFFIC-PDE-LEARN” in order to discover the hidden higher-order nonlinear PDE of complex traffic network. Specifically, the proposed model utilizes traffic measurements (i.e., density, flow, speed) collected from traffic sensors installed in highway networks. Since the traffic measurements collected from sensors are noisy and spatiotemporally sparse, the proposed model adopts multiple deep neural networks to reconstruct denoised and dense estimates of traffic measurements. Furthermore, the deep neural network can model the spatiotemporal heterogenous flow-density and flow-speed FDs existed in complex traffic network. Based on the denoised and dense estimates of traffic measurements, we explicitly design the proposed model structure in order to help the model discover the hidden PDE of the traffic network system. The derivation of higher-order nonlinear PDE models for complex traffic network is extremely challenging using the existing methods that are heavily based on first principles approach and prior knowledge of the systems, while the proposed “TRAFFIC-PDE-LEARN” model offers an alternative data-driven pathway that is able to unveil the intricate PDE model of complex traffic network, and keep interpretability.

\section{Problem Description and Framework}
The traffic network dynamics are considered to evolve on a spatiotemporal domain as defined by the Cartesian product $\Omega\times\Gamma$, where $\Omega\in\mathbb{R}^D$ is defined as the D-dimensional spatial domain and $\Gamma\in[t_0,t_m]$ is defined as the temporal domain, with $t_0$ and $t_m$ represent the starting and ending time steps. This study considers the spatial domain as 1-dimensional with $D=1$ that indicates the location along a highway and thus, the spatial domain can be expressed as $\Omega\in[x_0, x_m]$, where $x_0$ and $x_m$ denote the starting and ending locations. Traffic measurement variables including occupancy $o$, flow $q$, and speed $v$ evolve in the spatio-temporal domain $\Omega\times\Gamma$ as $o(x,t):[x_0,\ x_m]\times[t_0,t_m]\mapsto\mathbb{R}$, $q(x,t):[x_0,\ x_m]\times[t_0,t_m]\mapsto\mathbb{R}$, and $v(x,t):[x_0,\ x_m]\times[t_0,t_m]\mapsto\mathbb{R}$. Without prior knowledge regarding the traffic network dynamics, it is assumed that there exists a PDE model describing the dynamics, which can be formulated as the generic PDE form with a nonlinear $F$ as: 

\begin{equation} 
\label{hidden_PDE}
\partial_to=F\left(o,q,v,\partial_xo,\partial_xq,\partial_xv,\ldots,\partial_x^Mo,\partial_x^Mq,\partial_x^Mv\right),
\end{equation}

\noindent where $\partial_to=\frac{\partial o (x,t)}{\partial t}$ is the partial derivative of occupancy $o$ with respect to time step $t$ and it can be written as a function $F$ of $o(x,t)$, $q(x,t)$, $v(x,t)$ and their partial derivatives with respect to location $x$ up to the $M^{th}$ order. Specifically, $\partial_xo=\frac{\partial o(x,t)}{\partial x}$, $\partial_xq=\frac{\partial q(x,t)}{\partial x}$, and $\partial_xv=\frac{\partial v(x,t)}{\partial x}$ are defined as the first-order partial derivatives of $k$, $q$, $v$ with respect to $x$, while $\partial_x^Mo=\frac{\partial^Mo(x,t)}{\partial x^M}$, $\partial_x^Mq=\frac{\partial^Mq(x,t)}{\partial x^M}$, and $\partial_x^Mv=\frac{\partial^Mv(x,t)}{\partial x^M}$ denote their $M^{th}$ order partial derivatives with respect to $x$. For simplicity, the term $(x,t)$ is omitted in Eq.~\ref{hidden_PDE}.  The core objective of this study evolves learning $F$ to accurately describe the temporal evolution rate of traffic occupancy $\partial_to$.

To determine the unknown function $F$ in Eq.~\ref{hidden_PDE}, this study utilizes traffic measurement data (i.e., occupancy, flow, and speed) collected from traffic fixed-located sensors. By the nature of fixed-located sensors, the continuous spatiotemporal domain $\mathrm{\Omega}\times\mathrm{\Gamma}$ is further discretized into a two-dimensional grid comprising $m \times n$ points, where $m$ represents the number of spatial measurements and $n$ corresponds to the number of temporal measurements. The construction of this 2-D spatiotemporal grid assumes that sensors are evenly distributed along a highway at spatial intervals of $\Delta x$ and that data is collected at uniform temporal intervals of $\Delta t$. In practice, however, traffic sensors are typically installed at unequal spatial intervals, leading to missing traffic measurements at certain spatial locations on the 2-D grid. Furthermore, the data collected by these sensors is frequently affected by noise due to inherent sensor limitations. As a result, the traffic measurements obtained from the sensors can be expressed as:

\begin{equation}
\label{sparse_measurement}
    \tilde{o}(x,t), \quad \tilde{q}(x,t), \quad \tilde{v}(x,t), \quad for~ (x,t) \in O\backslash U, 
\end{equation}

\noindent where $O$ denotes the set of all $m\times n$ points on the spatiotemporal grid, while $U$ represents the set of missing points on the spatiotemporal grid. Variables $\tilde{o}$, $\tilde{q}$, and $\tilde{v}$ denote the noisy measurements of occupancy, flow, and speed, respectively, as observed from the traffic sensors.

\begin{figure}[!h] 
	\centering
	\includegraphics[width = \columnwidth]{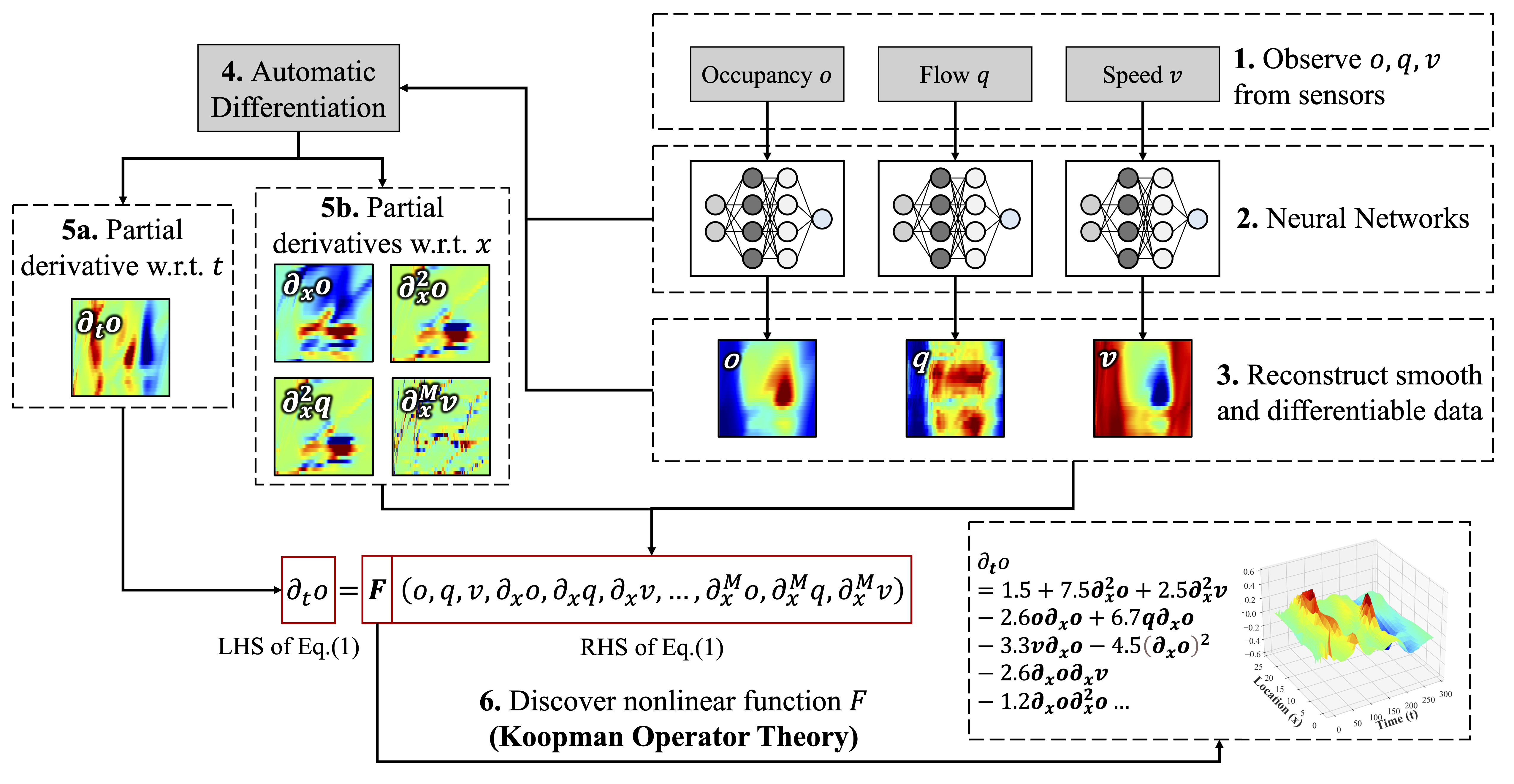}
	\caption{High-Level Design of the Proposed Model's Framework}
        \label{fig.framework}
\end{figure}

This study proposes the "TRAFFIC-PDE-LEARN" model to uncover the hidden partial differential equations (PDE) governing traffic network dynamics from noisy and sparse sensor measurements. The framework of the proposed model is illustrated in Fig.~\ref{fig.framework} and involves six key steps. In the first step, traffic measurement data are collected from sensors installed within the traffic network, which are inherently noisy and sparse due to sensor limitations. In the second step, the collected data are processed through separate neural networks, which reconstruct them and, with properly defined loss functions, result in smooth and differentiable traffic measurement data in the third step. Subsequently, in the fourth step, the proposed model computes the partial derivatives of $o$, $q$, and $v$ from the smooth and differentiable traffic data using Automatic Differentiation (Auto-Diff) applied to neural networks \cite{baydin2018automatic}. After obtaining the partial derivatives in the fifth step, the proposed model, in the sixth step, is explicitly designed to discover the unknown nonlinear function $F$ based on the reconstructed smooth and differentiable traffic data and their partial derivatives. The design in step 6 resonates with the framework applied in previous physical dynamics model discovery studies \cite{stephany2024pde}, to discover the nonlinear PDE enabled by Koopman theory, which offers a linear perspective on nonlinear dynamical systems by lifting the system into a higher-dimensional space of observable functions—functions of the system state. The detailed design of the proposed model is introduced in the methodology section.

\section{Methodology}
This section begins by detailing the proposed "TRAFFIC-PDE-LEARN" model components. Then it presents an overview of its holistic design, followed by a comprehensive explanation of its training procedure.

\subsection{Neural Networks Enabled Spatiotemporal Heterogeneous Fundamental Diagrams Approximation}

To begin with, since traffic occupancy $o$ is defined to evolve on $\Omega\times\Gamma$ as $o(x,t):\Omega\times\Gamma\mapsto\mathbb{R}$, it can be viewed as a function of $x$ and $t$ and the denoised and dense traffic occupancy can be reconstructed by a function $f_o$ as:

\begin{equation}
\label{Equ. Density Estimate}
    \hat{o}(x,t)=f_o(x,t), \quad for~(x,t)\in\Omega\times\Gamma, 
\end{equation}

\noindent where $\hat{o}$ denotes the reconstructed denoised and dense traffic occupancy and $f_o:\mathbb{R}^2\mapsto\mathbb{R}$ is a neural network approximated function. The relationship between $\hat{o}$ and the observed noisy and sparse occupancy from sensors $\widetilde{o}$ can be written as:  

\begin{equation}
\label{Equ. Noisy Estimate}
    \hat{o}(x,t)-\tilde{o}(x,t)=\epsilon_o(x,t), \quad for~(x,t)\in O\backslash U,
\end{equation}

\noindent where $\epsilon_o(x,t)$ denotes the random noise of observed occupancy at points with available data from sensors. Here $\tilde{o}$ is sparse with $(x,t)\in O\backslash U$, while $\hat{o}$ is dense with $(x,t)$ cover the entire spatiotemporal domain $\Omega\times\Gamma$. In order to obtain an accurate function $f_o$, the proposed model needs to find the $f_o$ that can minimize the difference between $f_o(x,t)$ and $\tilde{o}(x,t)$ during the training process, using a properly designed loss function, which will be introduced in section~\ref{holistic_design}.

\begin{figure}[!h] 
	\centering
	\includegraphics[width = \columnwidth]{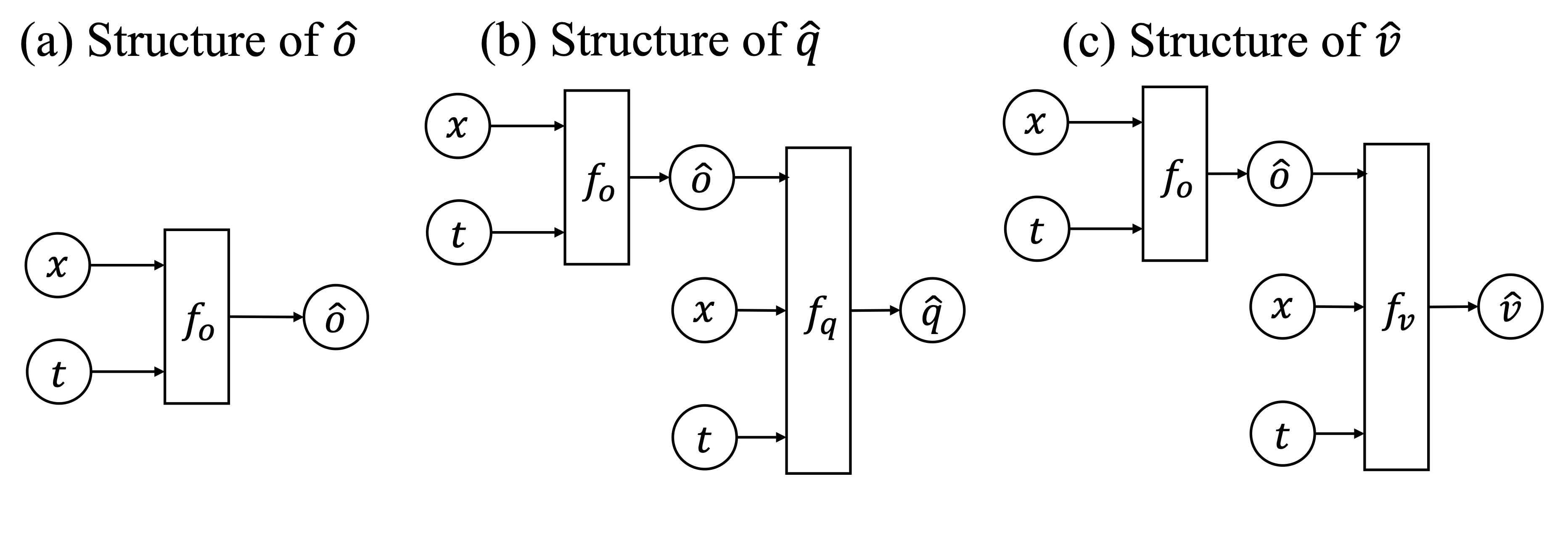}
	\caption{(a) Structure of $\hat{o}$, (b) Structure of $q$, and (c) Structure of $v$}
        \label{fig.structure_o_q_v}
\end{figure}

According to the flow-occupancy and speed-occupancy fundamental diagrams (FDs) \cite{geroliminis2011properties, wu2011empirical}, traffic flow can be expressed as a function of occupancy: $\hat{q} = Q(\hat{k})$, where $Q:\mathbb{R} \to \mathbb{R}$. Similarly, traffic speed can be represented as a function of occupancy: $\hat{v} = V(\hat{k})$, where $V:\mathbb{R} \to \mathbb{R}$. These relationships are referred to as homogeneous FDs, in which flow and speed depend solely on occupancy \cite{daganzo2006variational}. Consequently, the reconstructed denoised and dense traffic flow as well as speed for $(x,t)\in O\backslash U$, while $\hat{o}$,can be represented as:

\begin{equation}
\label{Equ. Flow Estimate FD}
    \hat{q}(x,t)=f_q(\hat{o},x,t)=f_q\left(f_o(x,t),x,t\right), 
\end{equation}

\begin{equation}
\label{Equ. Speed Estimate FD}
    \hat{v}(x,t)=f_v(\hat{o},x,t)=f_v\left(f_o(x,t),x,t\right), 
\end{equation}

\noindent where neural network approximated functions $f_q$ and $f_v$ become $f_q:\mathbb{R}^3\mapsto\mathbb{R}$ and $f_v:\mathbb{R}^3\mapsto\mathbb{R}$ and similarly, we have $\hat{q}(x,t)-\tilde{q}(x,t)=\epsilon_q(x,t)$ and $\hat{v}(x,t)-\tilde{v}(x,t)=\epsilon_v(x,t)$, for $(x,t)\in O\backslash U$, where $\epsilon_q$ and $\epsilon_v$ represent the random noise of flow and speed, respectively. In fact, Eq.~\ref{Equ. Flow Estimate FD} and Eq.~\ref{Equ. Speed Estimate FD} can also be referred as spatiotemporal heterogeneous FDs, where traffic flow and speed are not only functions of traffic occupancy but also of location and time. Different from homogenous FDs, Heterogenous FDs can represent the actual physical relationships between traffic measurements more realistically as these relationships can vary due to factors such as roadway geometric design, weather, and lighting conditions \cite{buisson2009exploring, seo2017traffic}. Similar to $f_o$, the proposed model obtains accurate functions $f_q$ and $f_v$ by minimizing the properly designed loss functions, which are introduced in section~\ref{holistic_design}, during the training process. The structures of $\hat{o}$, $\hat{q}$, and $\hat{v}$ are illustrated in Fig.~\ref{fig.structure_o_q_v}. 

Subsequently, functions $f_o$, $f_q$, and $f_v$ are approximated by designing them as three separate rational neural networks \cite{boulle2020rational}. Studies have proven that the structure of rational neural network is effective in approximating functions for discovering nonlinear PDEs from observations \cite{stephany2024pde, stephany2022pde}. For a neural network, at each of its hidden layer, a nonlinear activation function is applied to transform the inputs as $z=\sigma(Wi+b)$, where $i$ is the input vector, $z$ denotes the output vector of this hidden layer, $W$ is a trainable weight matrix, $b$ is the bias vector, and $\sigma$ represents the element-wise nonlinear activation function. The rational neural network applies rational functions defined in Eq.~\ref{Equ. rational function} as the activation function $\sigma$.

\begin{equation}\label{Equ. rational function}
    \sigma(m)=\frac{N(m)}{D(m)}=\frac{\sum_{i=0}^{r_N}{a_im^i}}{\sum_{j=0}^{r_D}{b_jm^j}},
\end{equation}

\noindent where $r_N$ and $r_D$ are the polynomial order of the numerator $N(m)$ and denominator $D(m)$, respectively, while $a_i$ and $b_i$ are trainable coefficients. The rational function $\sigma(m)$ is said to be the type of $(r_N,\ r_D)$. The major advantage of rational activation function is that they are more flexible and smoother, and have more function approximation power comparing with other types of activation functions such as ReLU and Sigmoid \cite{boulle2020rational}. 

\subsection{Computation of the Partial Derivatives of $\hat{o}$, $\hat{q}$, and $\hat{v}$}
Specifically, this section presents the computation processes of the partial derivatives up to the second order (i.e., $M=2$). Furthermore, for simplicity, the term $(x,t)$ is omitted from $\hat{o}(x,t)$, $\hat{q}(x,t)$, and $\hat{v}(x,t)$ in the following equations. For reconstructed traffic occupancy, $\hat{o}=f_o(x,t)$, since $f_o$ is a neural network approximated function, its first-order partial derivatives $\partial_t\hat{o}$ and $\partial_x\hat{o}$ can be easily derived through Auto-Diff \cite{baydin2018automatic}. Auto-Diff is a technique that leverages the chain rule to derive the exact derivatives of functions represented by computational graphs.

As for reconstructed traffic flow $\hat{q}=f_q(\hat{o},x,t)$, by introducing two intermediate variables $\phi\left(x\right)=x$ and $\lambda\left(t\right)=t$, $\hat{q}$ can be rewritten as $\hat{q}=f_q(\hat{o},\phi,\lambda)$. According to the chain rule, its first-order partial derivative with respective $x$ (i.e., $\partial_x\hat{q}$) can be derived as:

\begin{equation}
    \label{Equ. First Order 1}
    \partial_x\hat{q}=\partial_{\hat{o}}\hat{q}\partial_x\hat{o}+\partial_\phi\hat{q}\partial_x\phi+\partial_\lambda\hat{q}\partial_x\lambda.
\end{equation}

Based on the definition of intermediate variables $\phi$ and $\lambda$, it is obvious that $\partial_x\phi=1$ and $\partial_x\lambda=0$. Then, Eq.~\ref{Equ. First Order 1} becomes:

\begin{equation}\label{Equ. First Order 2}
\partial_x\hat{q}=\partial_{\hat{o}}\hat{q}\partial_x\hat{o}+\partial_\phi\hat{q},
\end{equation}

\noindent where $\partial_{\hat{o}}\hat{q}$, $\partial_x\hat{o}$, and $\partial_\phi\hat{q}$ can all be obtained easily through Auto-Diff. In a similar fashion, by rewriting $\hat{v}$ as $\hat{v}=f_v(\hat{o},\phi,\lambda)$, the first-order partial derivative of $\hat{v}$ with respective to $x$ (i.e., $\partial_x\hat{v}$) can be written as:

\begin{equation}\label{Equ. First Order 3}
    \partial_x\hat{v}=\partial_{\hat{o}}\hat{v}\partial_x\hat{o}+\partial_\phi\hat{v},
\end{equation}

\noindent where $\partial_{\hat{o}}\hat{v}$ and $\partial_\phi\hat{v}$ can also be obtained easily through Auto-Diff. 

Moving to the second-order partial derivatives, for reconstructed occupancy $\hat{o}=f_o(x,t)$, its second-order partial derivative with respect to x can be derived by applying Auto-Diff to $\partial_x\hat{o}$ without difficulties:

\begin{equation}\label{Equ. Second Order 1}
    \partial_x^2\hat{o}=\partial_x(\partial_x\hat{o}).
\end{equation}

For reconstructed flow $\hat{q}=f_q(\hat{o},x,t)$, the calculation of $\partial_x^2\hat{q}$ is conducted by applying the chain rule and product rule. Similar to the computation of $ \partial_x\hat{q}$, two intermediate variables are introduced (i.e., $\phi\left(x\right)=x$ and $\lambda(t)=t$) and rewrite $\hat{q}$ as $\hat{q}=f_q(\hat{o},\phi,\lambda)$. Its second-order partial derivative with respect to $x$ is calculated as follow:

\begin{equation}
\label{Equ. Second Order 2}
\begin{aligned}
\partial_x^2\hat{q}&=\partial_x(\partial_x\hat{q})\\
    &=\partial_x(\partial_{\hat{o}}\hat{q}\partial_x\hat{o}+\partial_\phi\hat{q}\partial_x\phi)=\partial_x(\partial_{\hat{o}}\hat{q}\partial_x\hat{o})+\partial_x(\partial_\phi\hat{q}\partial_x\phi)\\
    &=(\partial_x(\partial_{\hat{o}}\hat{q}))\partial_x\hat{o}+(\partial_x(\partial_x\hat{o}))\partial_{\hat{o}}\hat{q}\\
    &\quad+(\partial_x(\partial_\phi\hat{q}))\partial_x\phi+(\partial_x(\partial_x\phi))\partial_\phi\hat{q}\\
    &=\partial_{\hat{o}}^2\hat{q}(\partial_x\hat{o})^2+\partial_x^2\hat{o}\partial_{\hat{o}}\hat{q}+\partial_\phi^2\hat{q}(\partial_x\phi)^2+\partial_x^2\phi\partial_\phi\hat{q}\\&=\partial_{\hat{o}}^2\hat{q}(\partial_x\hat{o})^2+\partial_x^2\hat{o}\partial_{\hat{o}}\hat{q}+\partial_\phi^2\hat{q},
\end{aligned}
\end{equation}

\noindent where $\partial_{\hat{o}}\hat{q}$, $\partial_{\hat{o}}^2\hat{q}$, $\partial_x\hat{o}$, $\partial_x^2\hat{o}$, and $\partial_\phi^2\hat{q}$ can be obtained through Auto-Diff. By rewriting $\hat{v}$ as $\hat{v}=f_v(\hat{o},\phi,\lambda)$ and following the same process of Eq.~\ref{Equ. Second Order 2}, the second-order partial derivative of speed estimate $\hat{v}$ with respect to x is derived into the following:

\begin{equation}\label{Equ. Second Order 3}
    \partial_x^2\hat{v}=\partial_{\hat{o}}^2\hat{v}(\partial_x\hat{o})^2+\partial_x^2\hat{o}\partial_{\hat{o}}\hat{v}+\partial_\phi^2\hat{v},
\end{equation}

\noindent where $\partial_{\hat{o}}\hat{v}$, $\partial_{\hat{o}}^2\hat{v}$, and $\partial_\phi^2\hat{v}$ can also be obtained through Auto-Diff. Note that for higher order partial derivatives ($M>2$), they can be calculated by continuing to apply the chain rule and product rule along with Auto-Diff. 

\subsection{Hidden PDE Discovery of Traffic Network Dynamics by Koopman Theory}
Given the reconstructed denoised and dense traffic measurements $\hat{o}$, $\hat{q}$, $\hat{v}$, and their partial derivatives from the previous steps,  Eq.~\ref{hidden_PDE} can be rewritten into the following form:

\begin{equation}
\label{hidden_PDE_rewrite}
\partial_t\hat{o}=F(\hat{o},\hat{q},\hat{v},\partial_x\hat{o},\partial_x\hat{q},\partial_x\hat{v},\ldots,\partial_x^M\hat{o},\partial_x^M\hat{q},\partial_x^M\hat{v}).
\end{equation}

To discover $F$, the Koopman operator theory for PDEs \cite{kutz_applied_2018} framework is applied which provides a powerful linear representation of nonlinear dynamical systems by lifting them into a higher-dimensional function space. In this framework, instead of analyzing the nonlinear evolution of the system's state variables directly, one studies the evolution of a set of observable functions—functions $f_h$ defined on the state space—under a linear operator. This linearization in function space makes it possible to capture complex nonlinear behaviors using a weighted sum of basis observables, each of which evolves linearly over time. Such a perspective allows for systematic identification of governing equations from data, even when the original system dynamics are highly nonlinear. Further, computing higher-order derivatives requires more fine-grained data and hence we set $M=2$.  By that, the formulation  Eq.~\ref{hidden_PDE_rewrite} can be expressed as: 

\begin{equation}
\label{hidden_PDE_potential_terms_2_order}
\partial_t\hat{o}=\sum_{h=1}^{H}{p_hf_h(\hat{o},\hat{q},\hat{v},\partial_x\hat{o},\partial_x\hat{q},\partial_x\hat{v},\partial_x^2\hat{o},\partial_x^2\hat{q},\partial_x^2\hat{v})},
\end{equation}

\noindent where $H$ denotes the total number of potential observable functions, $f_h$ represents the $h^{th}$ potential observable function which is a scalar function with $\hat{o}$,$\hat{q}$,$\hat{v}$ and their partial derivatives up to the second order as inputs, and $p_h$ is the corresponding coefficient of the $h^{th}$ potential observable function $f_h$. Each potential observable function (any function that takes the state of the system as input and returns a value of interest, for example the product of the first-order derivatives of occupancy and speed) $h\in[1,H]$ is further defined to be monomial or equivalently, each $f_h$ is only made up by the power products of $\hat{o}$, $\hat{q}$, $\hat{v}$ and their partial derivatives. Thus, $f_h$ can be formulated into:

\begin{equation}
\label{potential_terms}
\begin{aligned}
&f_h(\hat{o},\hat{q},\hat{v},\partial_x\hat{o},\partial_x\hat{q},\partial_x\hat{v},\partial_x^2\hat{o},\partial_x^2\hat{q},\partial_x^2\hat{v})\\
&={\hat{o}}^{c_1(h)}{\hat{q}}^{c_2(h)}{\hat{v}}^{c_3(h)}(\partial_x\hat{o})^{c_4(h)}(\partial_x\hat{q})^{c_5(h)} \\
&\quad(\partial_x\hat{v})^{c_6(h)}(\partial_x^2\hat{o})^{c_7(h)}(\partial_x^2\hat{q})^{c_8(h)}(\partial_x^2\hat{v})^{c_9(h)},
\end{aligned}
\end{equation}

\noindent where $c_i(h)$, $i\in[1,\ldots,9]$ is the exponent of each variable in the monomial. With the definition of $f_h$ in Eq.~\ref{potential_terms}, the RHS of Eq.~\ref{hidden_PDE_potential_terms_2_order} becomes a polynomial of $\hat{o}$, $\hat{q}$, $\hat{v}$ and their partial derivatives. This polynomial is further defined to have an order of $N$ and this introduces a requirement for the exponents $c_i(h)$ in Eq.~\ref{potential_terms} to satisfy the following:

\begin{equation}
\label{potential_term_limitation}
\sum_{i=1}^{9}{c_i(h)\le N},\ \forall ~ h\in[1,\ldots,H].
\end{equation}

For the reconstructed denoised and dense traffic measurements defined in Eq.~\ref{Equ. Density Estimate}, Eq.~\ref{Equ. Flow Estimate FD}, and Eq.~\ref{Equ. Speed Estimate FD} and the hidden traffic network PDE model defined in Eq.~\ref{hidden_PDE_potential_terms_2_order}, the location and time variables $x$ and $t$ are defined on the spatiotemporal domain $\Omega\times\Gamma$ where $x$ and $t$ are continuous variables. However, for the convenience of designing the proposed model based on traffic measurements observed from traffic sensors , $x$ and $t$ should be considered as discrete variables. Thus, Eq.~\ref{hidden_PDE_potential_terms_2_order} can be reformulated into a vector format as the following: 

\begin{align}
\partial_t\hat{o}(x,t)&=\mathrm{\Theta}(x,t)^T\xi, \quad for~(x,t)\in O, \label{Equ. vector PDE}\\
\mathrm{\Theta}(x,t)&=[f_1(x,t),f_1(x,t),\ldots,f_H(x,t)]^T, \label{Equ. vector PDE Phi}\\
\xi&=[p_1,p_2,\ldots,p_H]^T, \label{Equ. vector PDE Xi}
\end{align}

\noindent where $\partial_t\hat{o}(x,t)\in\mathbb{R}$ is the time derivative of $\hat{o}$ at spatiotemporal point $(x,t)$. The location and time variable $x$ and $t$ are defined to evolve on a discrete spatiotemporal grid and $O$ denotes the collection of all points on this grid. $\mathrm{\Theta}(x,t)\in\mathbb{R}^H$ denotes a vector, where elements represent the $H$ number of potential function terms $f_h$ evaluated at point $(x,t)$. $\xi\in\mathbb{R}^H$ is a vector of the $H$ number of coefficients $p_h$ corresponding to potential function terms $f_h$. Note that the vector $\xi$ is a sparse vector, this can filter out the function terms $f_h$ that are unrelated to the hidden PDE, ensuring the discovered PDE is parsimonious via sparse regression.

\subsection{Holistic Design of the Proposed "TRAFFIC-PDE-LEARN" Model}
\label{holistic_design}
\begin{figure}[!h]
    \centering
    \includegraphics[width=\columnwidth]{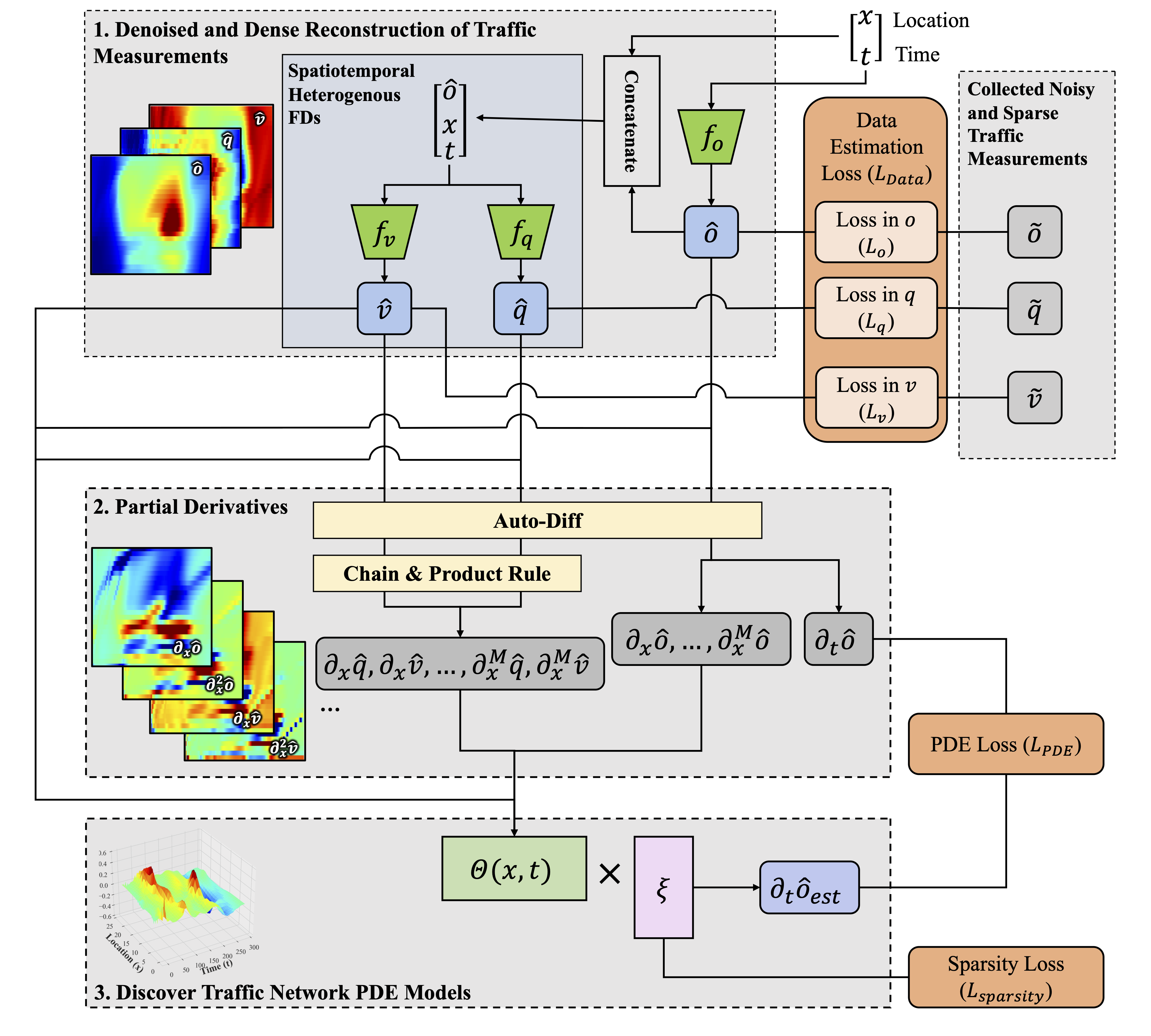}
    \caption{Holistic Design of the Proposed “TRAFFIC-PDE-LEARN” Model}
    \label{holisitic_model_design}
\end{figure}

The holistic design of the proposed model is illustrated in Fig.~\ref{holisitic_model_design}, which comprises three main components: (1) reconstruction of denoised and dense estimates of traffic measurements; (2) calculation of partial derivatives; and (3) discover the hidden PDEs of traffic network dynamics. The objectives of the proposed model are twofold: (1) find the neural network approximated functions $f_o$, $f_q$, and $f_v$, and (2) identify the sparse vector $\xi$. To fulfill these two objectives, multiple loss terms are explicitly designed to regulate the proposed model in order to correctly identify $f_o$, $f_q$, $f_v$, and $\xi$ during the model training process. The loss terms include data estimation loss ($L_{Data}$), PDE loss ($L_{PDE}$), and sparsity loss ($L_{sparsity}$). The data estimation loss  ($L_{Data}$) is defined as:

\begin{equation}
    \label{Eq.Loss_data}
    L_{Data}=\eta_oL_o+\eta_qL_q+\eta_vL_v,
\end{equation}

\noindent where $L_o$, $L_q$, and $L_v$ correspond to reconstruction loss in traffic occupancy, flow, and speed, respectively, calculated as the mean squared error (MSE) between the reconstructed denoised traffic measurement and the observed noisy traffic measurement from sensors. While, $\eta_o$, $\eta_q$, and $\eta_v$ denote the weights of $L_o$, $L_q$, and $L_v$, respectively. The formulations of $L_o$, $L_q$, and $L_v$ are as the following:

\begin{equation}
    \label{Eq.Loss_o}
    L_o=\frac{1}{N}\sum_{(x,t)~\in~O\backslash U}\left[\hat{o}(x,t)-\tilde{o}(x,t)\right]^2,
\end{equation}

\begin{equation}
    \label{Eq.Loss_q}
    L_q=\frac{1}{N}\sum_{(x,t)~\in~O\backslash U}\left[\hat{q}(x,t)-\tilde{q}(x,t)\right]^2,
\end{equation}

\begin{equation}
    \label{Eq.Loss_v}
    L_v=\frac{1}{N}\sum_{(x,t)~\in~O\backslash U}\left[\hat{v}(x,t)-\tilde{v}(x,t)\right]^2,
\end{equation}

\noindent where $N$ denotes the total number of points in set $O\backslash U$, while $\hat{o}$, $\hat{q}$, $\hat{v}$ represent the reconstructed denoised traffic measurements, and $\tilde{o}$, $\tilde{q}$, $\tilde{v}$ represent the corresponding observed noisy traffic measurements from sensors. Note that even though the reconstructed $\hat{o}$, $\hat{q}$, $\hat{v}$ are dense, which span through the entire spatiotemporal domain $\Omega\times\Gamma$, since the observed $\tilde{o}$, $\tilde{q}$, $\tilde{v}$ are sparse and only available at certain points on the spatiotemporal grid, $L_o$, $L_q$, and $L_v$ can only be calculated based on all points where traffic measurements are observable from sensors (i.e., $(x,t)\in O\backslash U$). 

The second loss term, PDE loss ($L_{PDE}$), is defined as the MSE between $\partial_t\hat{o}_{est}$ and $\partial_t\hat{o}$ as:

\begin{equation}
    \label{Eq.Loss_PDE}
    L_{PDE}=\frac{1}{N} \sum_{(x,t)~\in~O\backslash U}\left[\partial_t\hat{o}(x,t)-\partial_t\hat{o}_{est}\right]^2,
\end{equation}

\noindent where $\partial_t\hat{o}_{est} = \Theta(x,t)^T \xi$ is the time derivative of occupancy calculated via the discovered PDE models in the third component of Fig.~\ref{holisitic_model_design}, and $\partial_t\hat{k}$ is directly derived through auto-diff from the reconstructed denoised and dense occupancy $\hat{o}$. Similarly, $L_{PDE}$ is calculated based on all points where traffic measurements are observable from sensors (i.e., $(x,t)\in O\backslash U$).

The third loss term is the sparsity loss ($L_{spar}$) is calculated as the L1 norm of vector $\xi$. it is designed to regulate the proposed model to learn a sparse matrix $\xi$. This step is essential since it helps to filter out unrelated function terms of the hidden PDE and ensure the discovery of a parsimonious model: 

\begin{equation}
    \label{Eq.Loss_sparsity}
    L_{spar}= \left \Vert \xi \right \Vert_1.
\end{equation}

\subsection{Training Process of the Proposed Model}
The training process of the proposed "TRAFFIC-PDE-LEARN" model comprises three steps: the burn-in step, the main step, and the refinement step. During the initial burn-in step, the loss function to be minimized is defined in Eq.~\ref{Eq.Burn_in_loss}, where $L_{burn-in}$ consists of $L_{Data}$ and $L_{PDE}$. The purpose of this step is to guide the model in minimizing $L_{Data}$ to obtain accurate, denoised, and dense reconstructions of traffic measurements. Simultaneously, $L_{PDE}$ is included to regulate the model's discovery of the traffic dynamics PDE. However, because the reconstructed traffic measurements are not yet accurate during this step, the discovered PDE model is not expected to be precise at this stage. Instead, $L_{PDE}$ serves to ensure the smoothness of the training process.

\begin{align}
    L_{burn-in} & = L_{Data}+\eta_{PDE} L_{PDE} \nonumber\\
    & = \eta_{o} L_{o} + \eta_{q} L_{q} + \eta_{v} L_{v} + \eta_{PDE} L_{PDE} \label{Eq.Burn_in_loss}
\end{align}

After the initial burn-in step, as the denoised and dense reconstructions of traffic measurements become more accurate, the training process proceeds to the main step. In this step, the loss function $L_{main}$, defined in Eq.~\ref{Eq.main_loss}, is minimized. The objective is to further improve the denoised and dense reconstructions of traffic measurements while simultaneously discovering the correct PDE model. In addition to $L_{Data}$ and $L_{PDE}$, $L_{main}$ includes $L_{spar}$ to promote the sparsity of the vector $\xi$ defined in Eq.~\ref{Equ. vector PDE}. This sparsity ensures that the discovered PDE model is parsimonious. Specifically, we employ the sequential thresholding technique \cite{champion2019data} to enhance the sparsity of $\xi$. Thresholding is performed each time the process reaches a specific number of epochs during the main step. When thresholding is performed, the coefficients in $\xi$ that fall below a predefined threshold are set to zero for the remaining training epochs.

\begin{align}
    L_{main} & = L_{Data}+\eta_{PDE} L_{PDE} + \eta_{spar} L_{spar} \nonumber\\
    & = \eta_{o} L_{o} + \eta_{q} L_{q} + \eta_{v} L_{v} + \eta_{PDE} L_{PDE} + \eta_{spar} L_{spar} \label{Eq.main_loss}
\end{align}

The training process concludes with the refinement step, where the loss function $L_{refine}$, defined in Eq.~\ref{Eq.refinement_loss}, is minimized. $L_{refine}$ is structured similarly to $L_{burn-in}$, consisting only of $L_{Data}$ and $L_{PDE}$. During the refinement step, no thresholding is applied to $\xi$ to promote its sparsity. Instead, the training focuses on further refining the reconstructions of $\hat{o}$, $\hat{q}$, and $\hat{v}$, as well as the discovered PDE model.

\begin{align}
    L_{refine} & = L_{Data}+\eta_{PDE} L_{PDE} \nonumber\\
    & = \eta_{o} L_{o} + \eta_{q} L_{q} + \eta_{v} L_{v} + \eta_{PDE} L_{PDE} \label{Eq.refinement_loss}
\end{align}

Note that when selecting the weights of each loss term (i.e., $\eta_{o}$, $\eta_{q}$, $\eta_{v}$, $\eta_{PDE}$, and $\eta_{spar}$), it is ideal to adjust them so that the magnitudes of all loss terms are at the same level. However, if the training process needs to prioritize a specific loss term, its corresponding weight can be increased.

\section{Experimental Results}
\subsection{Data Preparation}
\begin{figure}[!h] 
	\centering
	\includegraphics[width = \columnwidth]{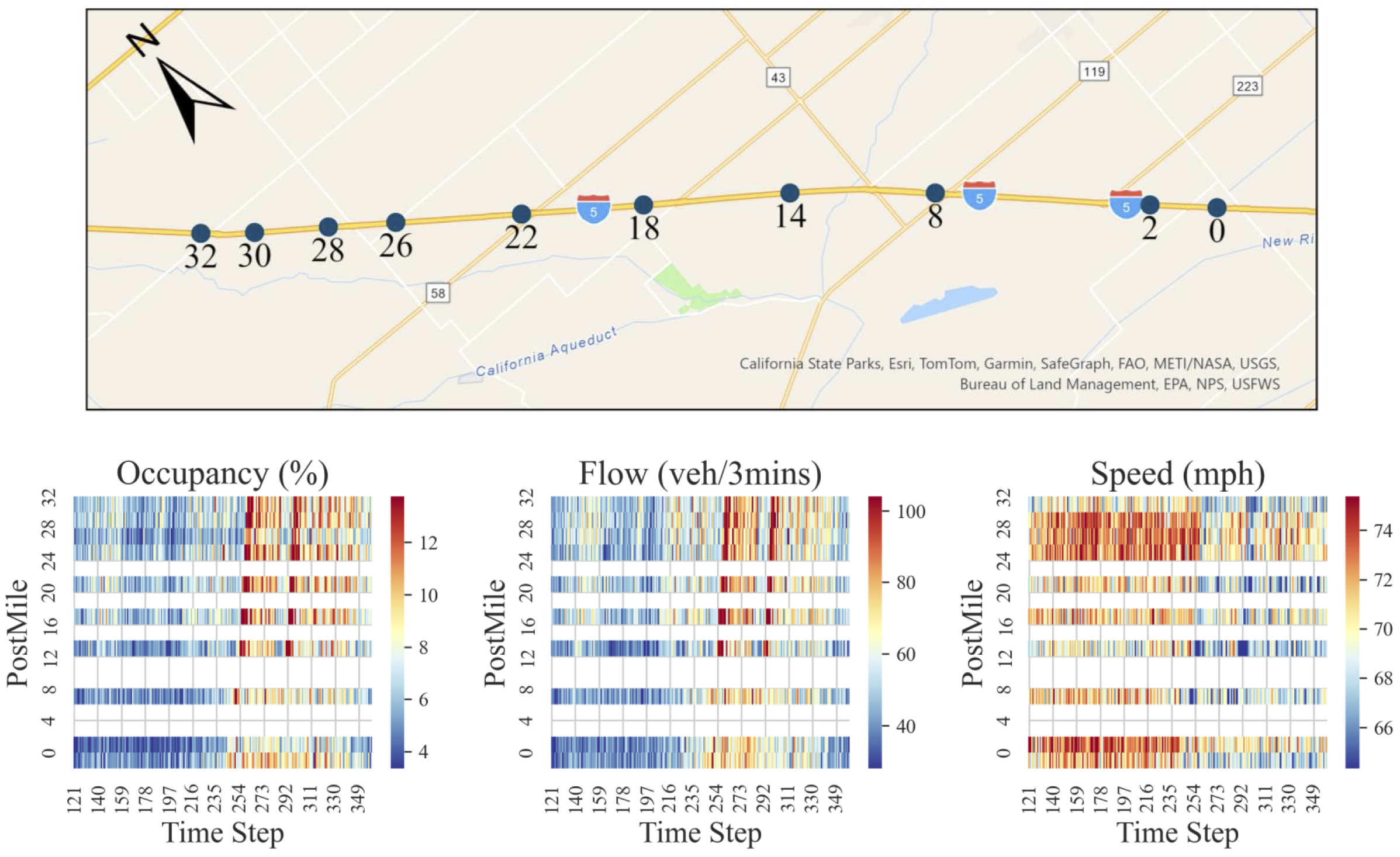}
	\caption{Locations of the Traffic Sensors, and the Corresponding Noisy and Sparse Traffic Measurement}
        \label{Fig. PDE Data}
\end{figure}

The traffic measurement data utilized in this study, including occupancy, flow, and speed, are collected from the Performance Measurement System (PeMS) of the California Department of Transportation (Caltrans). Traffic measurement data are collected from a 32-mile-long section of Interstate 5 (I-5) North direction in Kern County, California (see Fig.~\ref{Fig. PDE Data}). The PDE model's temporal step size $\partial t$ is set to 3 minutes in this study. The raw traffic measurements data are collected from PeMS with a 30-second temporal interval and are aggregated into 3-minute interval. Data are collected for a 1-day period from 00:00 to 23:55 on August $15^{\text{th}}$, 2023, with a 3-minute interval. This results in a total number of 480 time steps. However, since traffic volumes during the early morning and late night are relatively low, we only include traffic measurement data from 06:00 to 18:00 (i.e., time steps 120 to 360) into the analysis. Furthermore, the spatial step size $\partial x$ is set to 2 miles, so the 32-mile-long section can be segmented into 16 subsections with 17 sensor locations. In the PeMS database, traffic sensors are only available at 10 out of 17 locations on this 32-mile-long highway section. The locations with available traffic measurement data are marked in the upper figure of Fig.~\ref{Fig. PDE Data}, and the heatmaps of the collected traffic measurement data are presented in the lower plot of Fig.~\ref{Fig. PDE Data}.
 
\subsection{Training}
The training process of the proposed model is presented in this section. To begin with, the proposed model’s hyperparameters are set as the values listed in Table~\ref{Tab.Hyperparameter_value}. This settings of the hyperparameters are determined after extensive random search, However, given the large scale of the proposed model, exploring all possible combinations would inevitably require a significant amount of computational resources. Therefore, the hyperparameters are tuned within a limited search space.The detailed hyperparameters are summarized in Table I. 

\begin{table}[!h]
\centering
\scriptsize
\caption{Hyperparameters of the Proposed ``TRAFFIC-PDE-LEARN'' Model}
\label{Tab.Hyperparameter_value}
\begin{tabular}{lclc}
\hline
\textbf{Hyperparam.} & \textbf{Value} & \textbf{Hyperparam.} & \textbf{Value} \\ \hline
1. Learning Rate & 0.001 & 11. $\eta_v$ & 0.6 \\
2. LR Shrink Rate & 0.9 & 12. $\eta_{PDE}$ & 10 \\
3. LR Step Size & 500 & 13. $\eta_{spar}$ & 0.1 \\
4. Width $f_k$ & [2,50,100,100,50,1] & 14. PDE Order $M$ & 2 \\
5. Width $f_q$ & [3,50,100,100,50,1] & 15. Poly Order $N$ & 2 \\
6. Width $f_v$ & [3,50,100,100,50,1] & 16. Coeff. Thresh. & 0.0005 \\
7. Poly Ord. $r_N$ & 3 & 17. Thresh. Freq. & 400 \\
8. Poly Ord. $r_D$ & 2 & 18. Burn-in Epochs & 1000 \\
9. $\eta_k$ & 2 & 19. Main Epochs & 2500 \\
10. $\eta_q$ & 0.04 & 20. Refine Epochs & 1500 \\ \hline
\end{tabular}
\end{table}

\begin{figure}[!h]
    \centering
    \includegraphics[width=\columnwidth]{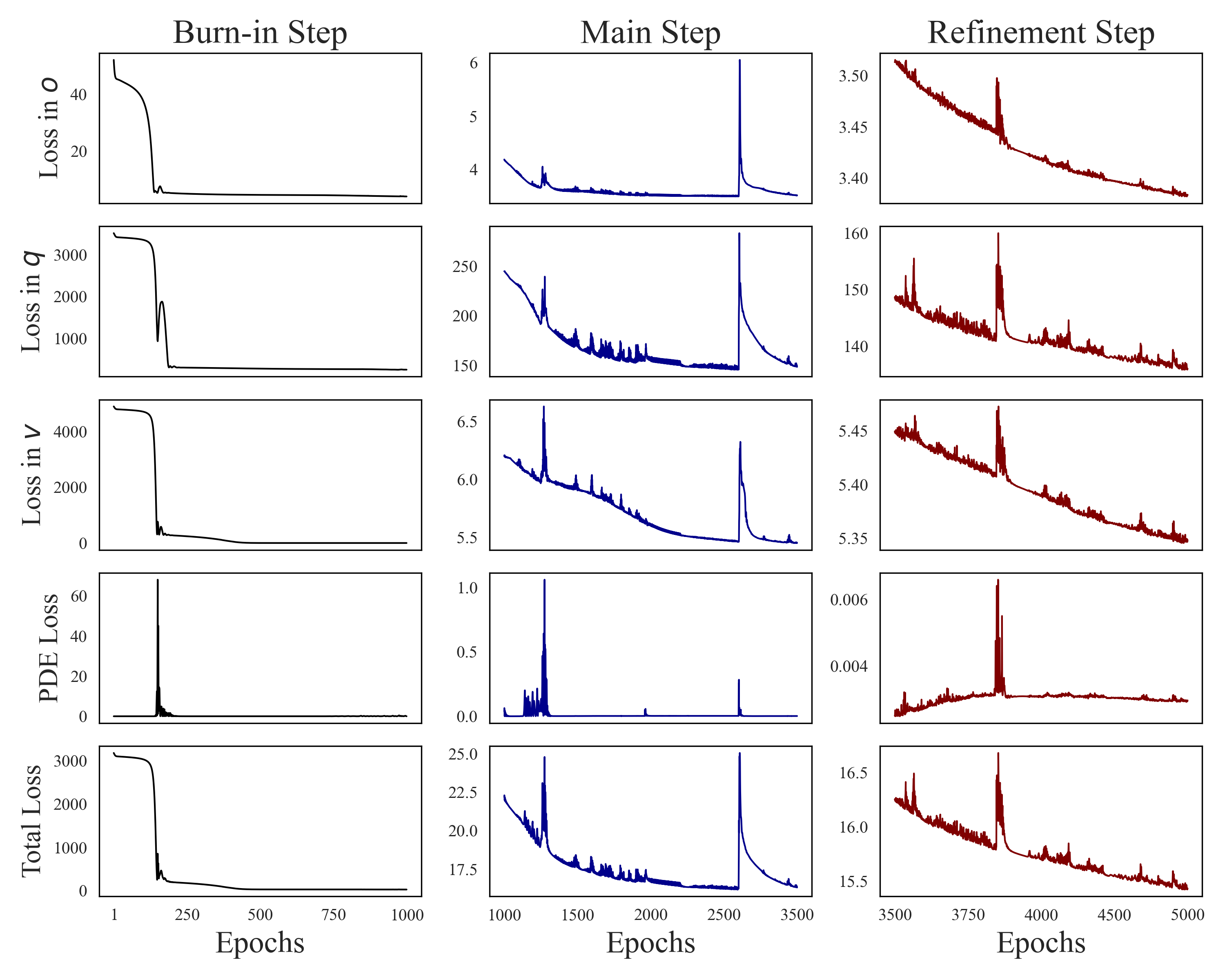}
    \caption{Loss Functions' Values during Burn-in, Main, and Refinement Steps.}
    \label{Fig.Loss}
\end{figure}

Subsequently, the loss curves of each loss function during the training process are presented in Fig.~\ref{Fig.Loss}. The training process consists of three steps: the burn-in step, the main step, and the refinement step. During the initial burn-in step, the values of the loss in $o$ ($L_o$), loss in $q$ ($L_q$), loss in $v$ ($L_v$), and the burn-in total loss ($L_{burn-in}$) decrease significantly. As for the PDE loss ($L_{PDE}$), during the initial burn-in step, it appears to be unstable with a number of large loss spikes. The appearance of an excessive number of large loss spikes indicates that the model did not discover the correct coefficients in vector $\xi$. When the model attempts to discover the correct coefficients in vector $\xi$, it simultaneously constructs denoised and dense traffic measurements. Any changes in them will lead to sensitive changes of $\partial_t\hat{o}$. As the training process continues during the burn-in step, both the intensity and magnitude of the loss spikes in $L_{PDE}$ become smaller. This indicates that the coefficients in vector $\xi$ become closer to the desired values. During the main step, $L_o$, $L_q$, $L_v$, and the main step total loss ($L_{main}$) continue to decrease. For $L_{PDE}$, several large loss spikes can still be observed; however, the magnitudes of these loss spikes are much smaller than those in the burn-in step. During the final refinement step, $L_o$, $L_q$, $L_v$, and the refinement step total loss ($L_{refine}$) continue to decrease steadily without the existence of large loss spikes. For $L_{PDE}$, the evolution of its value becomes stable and remains at a lower level during the refinement step. 

\begin{figure}[!h] 
	\centering
	\includegraphics[width = \columnwidth]{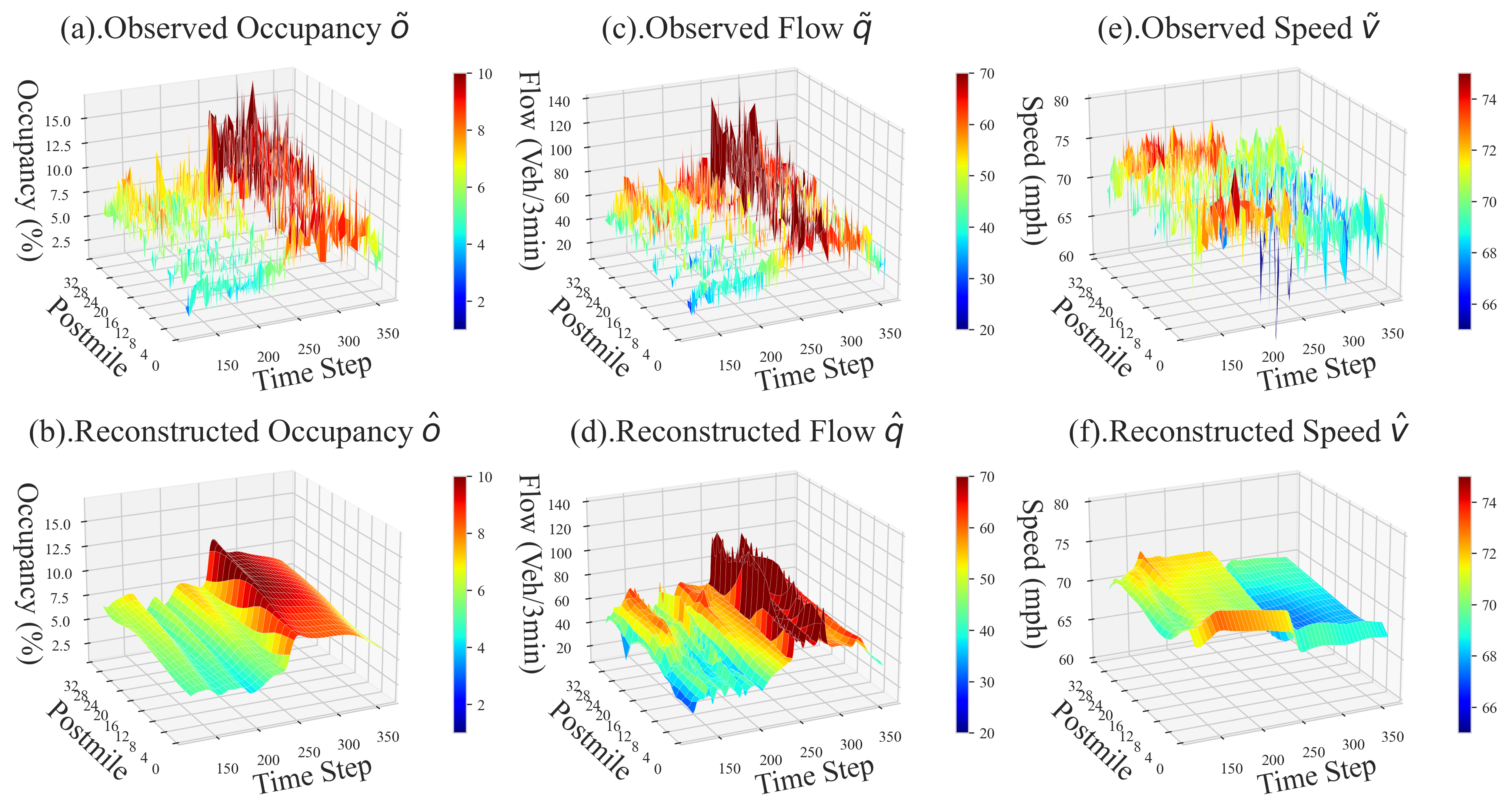}
	\caption{Observation v.s. Reconstruction}
        \label{Fig. Recon Surface Plots}
\end{figure}

The observed noisy and sparse traffic occupancy, flow, and speed (i.e., $\tilde{o}$, $\tilde{q}$, $\tilde{v}$) and their reconstructed counterparts (i.e., $\hat{o}$, $\hat{q}$, and $\hat{v}$) are presented in Fig~\ref{Fig. Recon Surface Plots}. In particular, Fig.~\ref{Fig. Recon Surface Plots} (a), (c) and (e) present the observed traffic occupancy, flow, and speed measurements, respectively, and Fig.~\ref{Fig. Recon Surface Plots} (b), (d) and (f) present the reconstructed traffic occupancy, flow, and speed, respectively. The observed traffic measurement data are missing on multiple spatiotemporal points and the data contains noises as the data exhibit excessive numbers of fluctuations. The proposed "TRAFFIC-PDE-FIND" model reconstructs the traffic measurement data through neural network approximated functions $f_o(x,t)$ (Eq.~\ref{Equ. Density Estimate}), $f_q(\hat{k},x,t)$ (Eq.~\ref{Equ. Flow Estimate FD}), and $f_v(\hat{k},x,t)$ (Eq.~\ref{Equ. Speed Estimate FD}). The reconstructed traffic measurements $\hat{o}$, $\hat{q}$, and $\hat{v}$ are smoother and cover all the missing spatiotemporal points. This indicates that $f_o(x,t)$,  $f_q(x,t)$, and  $f_v(x,t)$ of the proposed "TRAFFIC-PDE-LEARN" model are capable of learning the hidden spatiotemporal patterns and relationships of the traffic measurements and reconstruct reliable denoised and dense estimates of them. The reconstruction root mean squared error (RMSE) of traffic occupancy, flow, and speed are 1.84\%, 11.67veh/3mins, and 2.31mph, respectively. The RMSE is defined as:

\begin{equation}\label{Equ. PDE Recon RMSE}
    RMSE=\sqrt{\frac{1}{|O\backslash U|}\sum_{(x,t)~\in~O\backslash U}{(\hat{m}(x,t)-\tilde{m}(x,t))}},
\end{equation}


\noindent where $O\backslash U$ is the set of points with observed traffic measurement data, $|O\backslash U|$ denotes the total number of points in set $O\backslash U$, $\hat{m}$ is the reconstructed traffic measurement (i.e., $\hat{o}$, $\hat{q}$, $\hat{v}$), and $\tilde{m}$ represents the observed traffic measurements (i.e., $\tilde{o}$, $\tilde{q}$, $\tilde{v}$). Subsequently, the discovery of the traffic network dynamics' hidden PDE by the proposed model is based on the reconstructed traffic measurements $\hat{o}$, $\hat{q}$, $\hat{v}$.

\begin{figure}[!h] 
	\centering
	\includegraphics[width = \columnwidth]{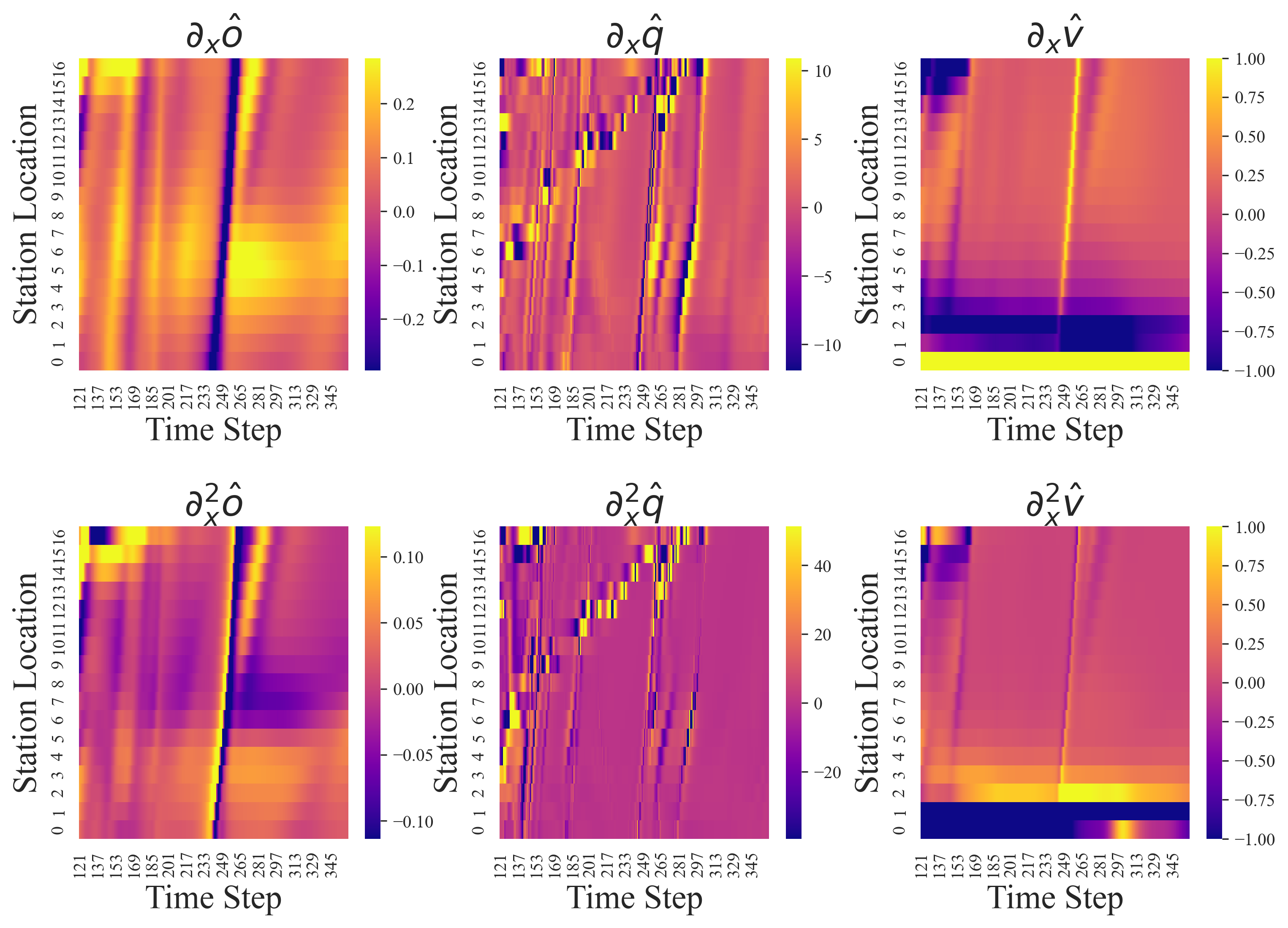}
	\caption{Partial Derivatives with Respect to Location x of the denoised and Dense Estimates of Traffic Measurements (i.e., inputs of the function F on RHS of Eq.~\ref{hidden_PDE_rewrite})}
        \label{Fig. Partial Derivatives}
\end{figure}

After obtaining the reconstructed denoised and dense traffic measurements $\hat{o}$, $\hat{q}$, and $\hat{v}$, their partial derivatives with respect to location x are derived through Auto-Diff plus the chain rule and product rule as defined from Eq.~\ref{Equ. First Order 1} to \ref{Equ. Second Order 3}. These partial derivatives (i.e. $\partial_x\hat{o}$, $\partial_x\hat{q}$, $\partial_x\hat{v}$, $\partial_x^2\hat{o}$, $\partial_x^2\hat{q}$, $\partial_x^2\hat{v}$) are presented in Fig.~\ref{Fig. Partial Derivatives}. Along with $\hat{o}$, $\hat{q}$ and $\hat{v}$, they together serve as the inputs of the unknown function $F$ on the RHS of the hidden PDE equation defined in Eq.~\ref{hidden_PDE_rewrite}. 

The LHS of Eq.~\ref{hidden_PDE_rewrite}, $\partial_t\hat{o}$, is derived through applying Auto-Diff in the "TRAFFIC-PDE-FIND" model. Subsequently, "TRAFFIC-PDE-FIND" discovers the traffic PDE model in Eq.~\ref{hidden_PDE_rewrite} by identifying the suitable sparse vector $\xi$ defined in Eq.~\ref{Equ. vector PDE Xi}. The discovered traffic network dynamics PDE model is:

{\small
\begin{equation}\label{Equ. Discovered PDE}
    \begin{aligned}
    \partial_t o &= 5.7 \times 10^{-2} - 1.4 \times 10^{-2} \partial_x o - 1.6 \times 10^{-2} \partial_x v - 1.0 \times \\&10^{-2} \partial_x^2 o + 2.3 \times 10^{-2} \partial_x^2 v - 4.3 \times 10^{-2} o \partial_x o 
    - 5.6 \times 10^{-3} o \\&\partial_x v - 5.6 \times 10^{-3} q \partial_x o+ 3.5 \times 10^{-4} q \partial_x^2 v 
    + 2.6 \times 10^{-1} (\partial_x o)^2 - \\&3.7 \times 10^{-2} \partial_x o \partial_x v - 8.9 \times 10^{-2}\partial_x o \partial_x^2 o + 1.0 \times 10^{-5} \partial_x o \partial_x^2 q - \\&1.2 \times 10^{-2} \partial_x o \partial_x^2 v -4.7 \times 10^{-4} q_x \partial_x v + 4.7 \times 10^{-4} q_x \partial_x^2 o\\& - 2.3 \times 10^{-4} q_x \partial_x^2 v + 1.1 \times 10^{-2} (\partial_x v)^2 -3.6 \times 10^{-2} \partial_x v \partial_x^2 o\\& + 7.0 \times 10^{-4} \partial_x v \partial_x^2 v - 3.2 \times 10^{-2} (\partial_x^2 o)^2 - 4.0 \times 10^{-5} \partial_x^2 o \partial_x^2 q\\& + 2.6 \times 10^{-3} \partial_x^2 o \partial_x^2 v + 3.7 \times 10^{-3} (\partial_x^2 v)^2,
     \end{aligned}
\end{equation}
}

\begin{figure}[!h] 
	\centering
	\includegraphics[width = \columnwidth]{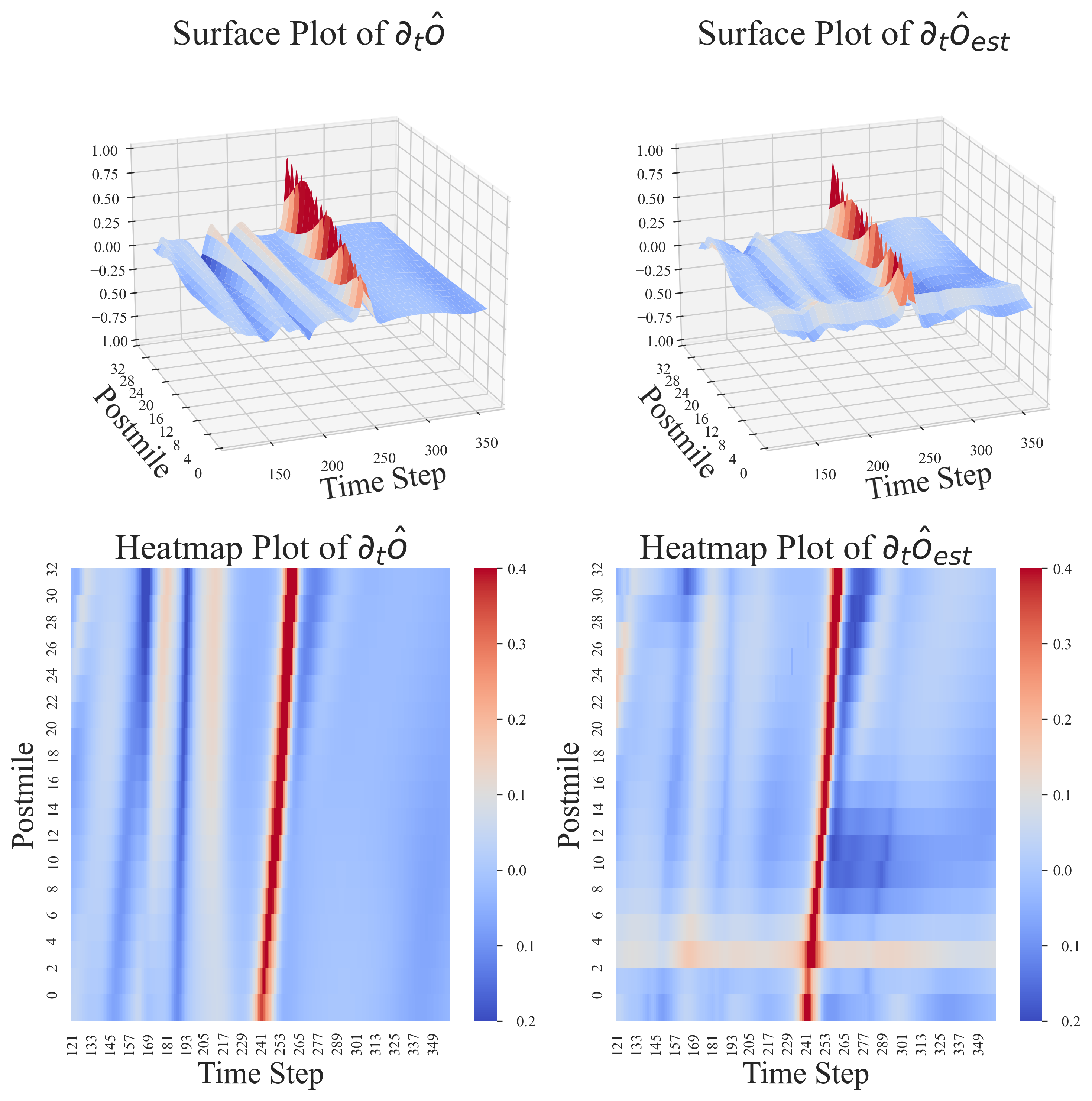}
	\caption{Comparison between $\partial_t\hat{k}$ Derived from Auto-Diff and $\partial_t{\hat{k}}_{est}$ Calculated through the Discovered PDE Model}
        \label{Fig. Discovered PDE Comparsion}
\end{figure}

\begin{figure}[!h] 
	\centering
	\includegraphics[width = 0.8\columnwidth]{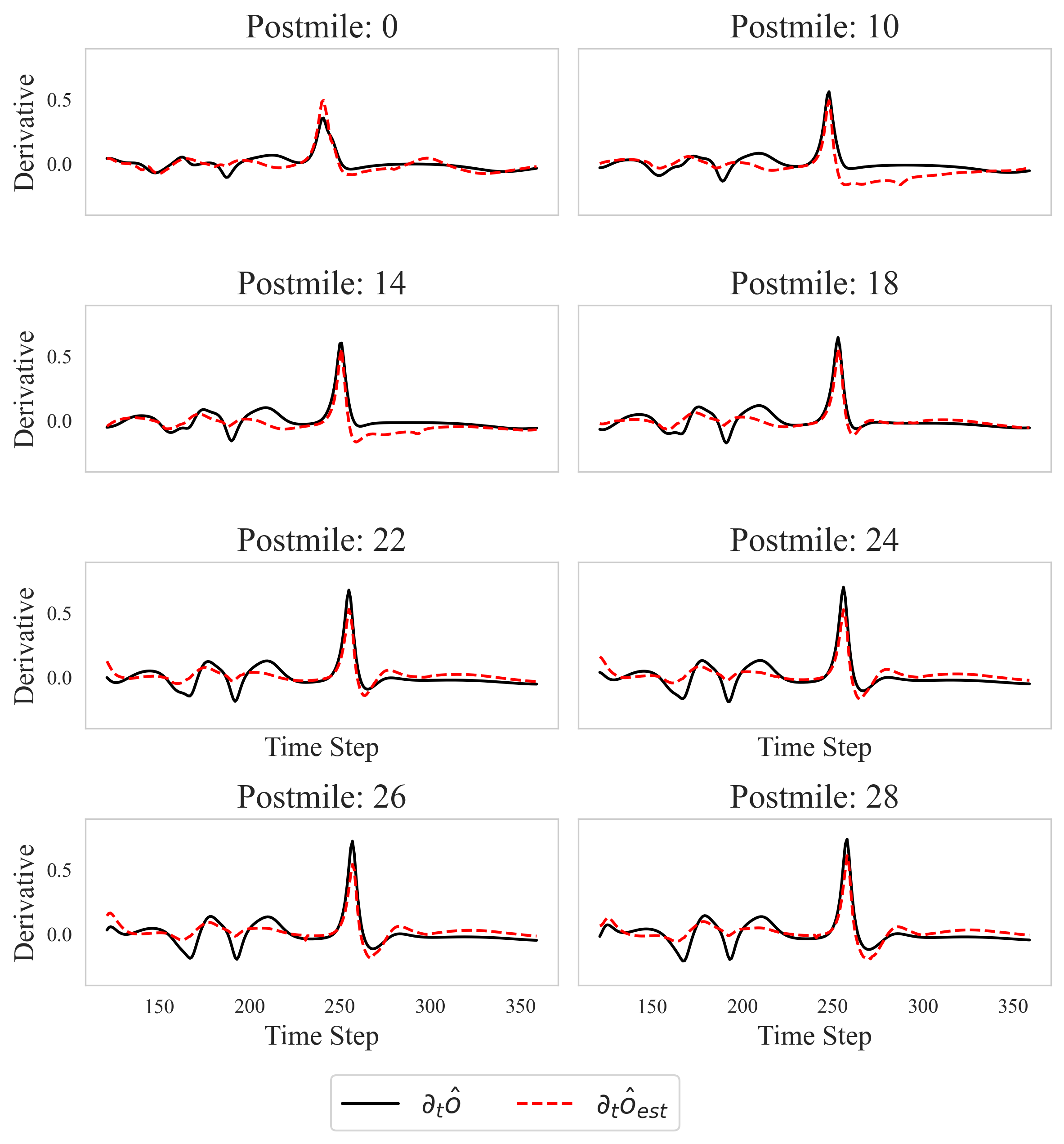}
	\caption{Discovered PDE Dynamics Compared with Actual Dynamics}
        \label{Fig. Discovered PDE Comparsion Lines}
\end{figure}

\noindent The time derivative of traffic occupancy $o$ calculated via the discovered PDE model in Eq.~\ref{Equ. Discovered PDE} is denoted as $\partial_t{\hat{o}}_{est}$. In Fig.~\ref{Fig. Discovered PDE Comparsion}, the comparison between $\partial_t\hat{o}$ and $\partial_t{\hat{o}}_{est}$ is presented. It can be observed that multiple key features of $\partial_t\hat{o}$ are captured in the spatiotemporal domain by the discovered PDE model. Furthermore, in Fig.~\ref{Fig. Discovered PDE Comparsion Lines}, we compare $\partial_t\hat{o}$ and $\partial_t{\hat{o}}_{est}$ for eight selected locations in the network as time-series plots, which also indicates the accuracy of the discovered PDE model in Eq.~\ref{Equ. Discovered PDE}, demonstrating the efficacy of the proposed “TRAFFIC-PDE-LEARN” model in discovering reliable traffic network dynamics PDE models.

\subsection{Traffic Flow Prediction based on the Discovered PDE Model}
\begin{figure}[!h] 
	\centering
	\includegraphics[width = \columnwidth]{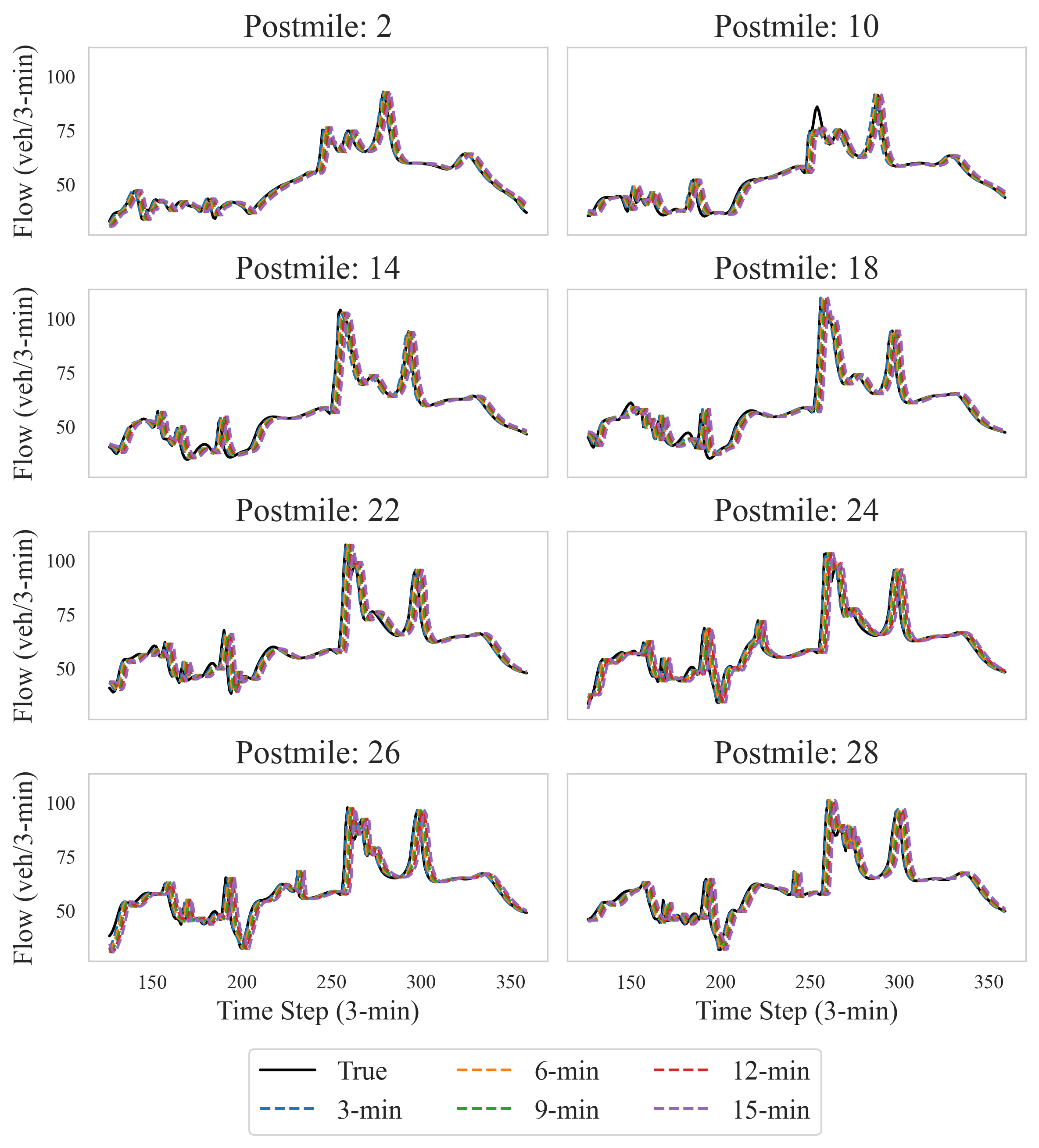}
	\caption{Traffic Flow Predictions based on the Discovered PDE Dynamics Model}
        \label{Fig. Flow Prediction}
\end{figure}

To further demonstrate the accuracy of the discovered PDE dynamics model in Eq.~\ref{Equ. Discovered PDE}, the model is employed to make forward predictions of traffic flow $q$. Specifically, this section presents one-step (3-minute) to five-step (15-minute) predictions of traffic flow. To predict traffic flow over multiple steps, we first predict traffic occupancy $\hat{o}$ by applying the forward Euler method as:

\begin{equation}\label{Equ. Euler Method}
    \hat{o}(x,t+1) = \hat{o}(x,t) + \partial_t\hat{o}(x,t),
\end{equation}

\noindent and then apply the flow-occupancy FD: $\hat{q}(x,t)=f_q(\hat{o},x,t)$, discovered by the proposed model, to obtain the traffic flow. The time derivative of occupancy: $\partial_t\hat{o}(x,t)$ in Eq.~\ref{Equ. Euler Method} is computed by Eq.~\ref{Equ. Discovered PDE}, in which $\hat{q}(x,t)$ is equal to $f_q(\hat{o},x,t)$, $\hat{v}(x,t)$ is equal to $f_v(\hat{o},x,t)$, and their partial derivatives with respect to $x$ are calculated using numerical differentiation. Fig.~\ref{Fig. Flow Prediction} presents the traffic flow prediction results in comparison with the true traffic flow values.

Additionally, we compare the prediction performance of the proposed model (using Eq.\ref{Equ. Discovered PDE}) with other existing traffic dynamics models, including the CTM model \cite{daganzo1994cell, daganzo1995cell}, and two AI-based time-series models: long short-term memory (LSTM) and recurrent neural network (RNN). To further highlight that traffic network dynamics PDE models are inherently higher-order, we modify the proposed "TRAFFIC-PDE-LEARN" model to discover a first-order nonlinear PDE model and compare its prediction performance with that of the proposed model in Eq.\ref{Equ. Discovered PDE}.

The traffic flow prediction results ranging from 3-minute to 15-minute are compared in Table~\ref{table:prediction_results_comparison}. Across all prediction horizons (3-min to 15-min), the higher-order nonlinear PDE model discovered by the proposed "TRAFFIC-PDE-LEARN" model consistently outperforms all other models, achieving the lowest RMSE and MAPE values. In particular, it significantly improves prediction accuracy compared to the CTM, and AI-based models (LSTM and RNN). Even though the data-driven first-order nonlinear PDE model improves the prediction accuracy compared with the existing models, higher-order still performs best, highlighting the superior predictive capability and robustness of the proposed model across different forecasting intervals. 

\begin{table}[!h]
\centering
\scriptsize
\caption{Traffic Flow Prediction Based on Different Models (Unit: veh/3-min)}
\begin{tabular}{ccccccc}
\hline
Model & Metric & 3-min & 6-min & 9-min & 12-min & 15-min \\ \hline
\multirow{2}{*}{CTM} & RMSE & 5.57 & 7.66 & 8.81 & 9.39 & 10.00 \\
                     & MAPE & 4.99 & 7.44 & 8.83 & 9.96 & 10.73 \\ \cline{2-7}
\multirow{2}{*}{LSTM} & RMSE & 8.74 & 9.02 & 9.54 & 10.41 & 11.69 \\
                      & MAPE & 11.55 & 12.10 & 12.85 & 13.84 & 15.17 \\ \cline{2-7}
\multirow{2}{*}{RNN} & RMSE & 8.00 & 8.95 & 10.24 & 11.64 & 13.01 \\
                     & MAPE & 9.87 & 11.23 & 12.99 & 14.65 & 16.18 \\ \cline{2-7}
\multirow{2}{*}{DD-PDE} & RMSE & 3.66 & 5.18 & 7.08 & 8.88 & 10.36 \\
                        & MAPE & 2.54 & 4.03 & 5.63 & 7.16 & 8.52 \\ \cline{2-7}
\multirow{2}{*}{\textbf{TPDE (Ours)}} & RMSE & \textbf{1.58} & \textbf{3.00} & \textbf{4.73} & \textbf{6.18} & \textbf{7.32} \\
                                      & MAPE & \textbf{1.59} & \textbf{3.06} & \textbf{4.81} & \textbf{6.35} & \textbf{7.66} \\ \hline
\end{tabular}
\label{table:prediction_results_comparison}
\end{table}


\section{Conclusion}
The modeling of traffic network dynamics is of pivotal importance for facilitating the prediction, management, and control of traffic network states. Previous efforts have extensively developed PDEs for traffic network dynamics by mathematically investigating the physical laws of traffic networks, aiming to derive governing equations from first principles. However, these PDE models often fail to accurately capture real-world network dynamics due to the complexity of traffic networks, which frequently involve higher-order nonlinearities that make deriving first principles mathematically challenging. Recent advancements in physics-informed artificial intelligence and the abundance of data collected from real-world physical systems have shifted the focus of developing governing laws from first-principles methods to data-driven approaches. In this study, we propose a novel data-driven deep learning model, "TRAFFIC-PDE-LEARN," capable of uncovering hidden traffic network PDEs from sparse and noisy traffic measurements. Specifically, our proposed model operates in three stages. First, it leverages neural network approximations based on spatiotemporal heterogeneous FDs to reconstruct denoised and dense traffic measurements. Second, it derives the partial derivatives of traffic measurements using Auto-Diff, the chain rule, and the product rule. Third, based on the Koopman operator theory for PDE, it formulates the hidden PDE model by including multiple potential function terms and identifies the active terms along with their corresponding coefficients. The proposed model is trained using real-world traffic network data observed through traffic sensors. Notably, the model is designed to discover a second-order nonlinear PDE model of traffic network dynamics. The results demonstrate that "TRAFFIC-PDE-LEARN" successfully discovers accurate higher-order nonlinear PDE models of traffic network dynamics. Furthermore, the discovered PDE model is applied to predict traffic flow over time horizons ranging from 3 to 15 minutes. The predictions are compared with state of art models. The results show that the proposed model outperform the others. 

This study has several limitations and directions for future research. The unequal spacing of traffic sensors in PeMS contributes to the sparsity of traffic measurements, potentially leading to information loss. Future studies could explore the use of connected automated vehicles to collect more fine-grained data to further validate the framework.

\bibliographystyle{IEEEtran}
\bibliography{references}

\vspace{-1.5em}
\begin{IEEEbiography}[{\includegraphics[width=1in,height=1.25in,clip,keepaspectratio]{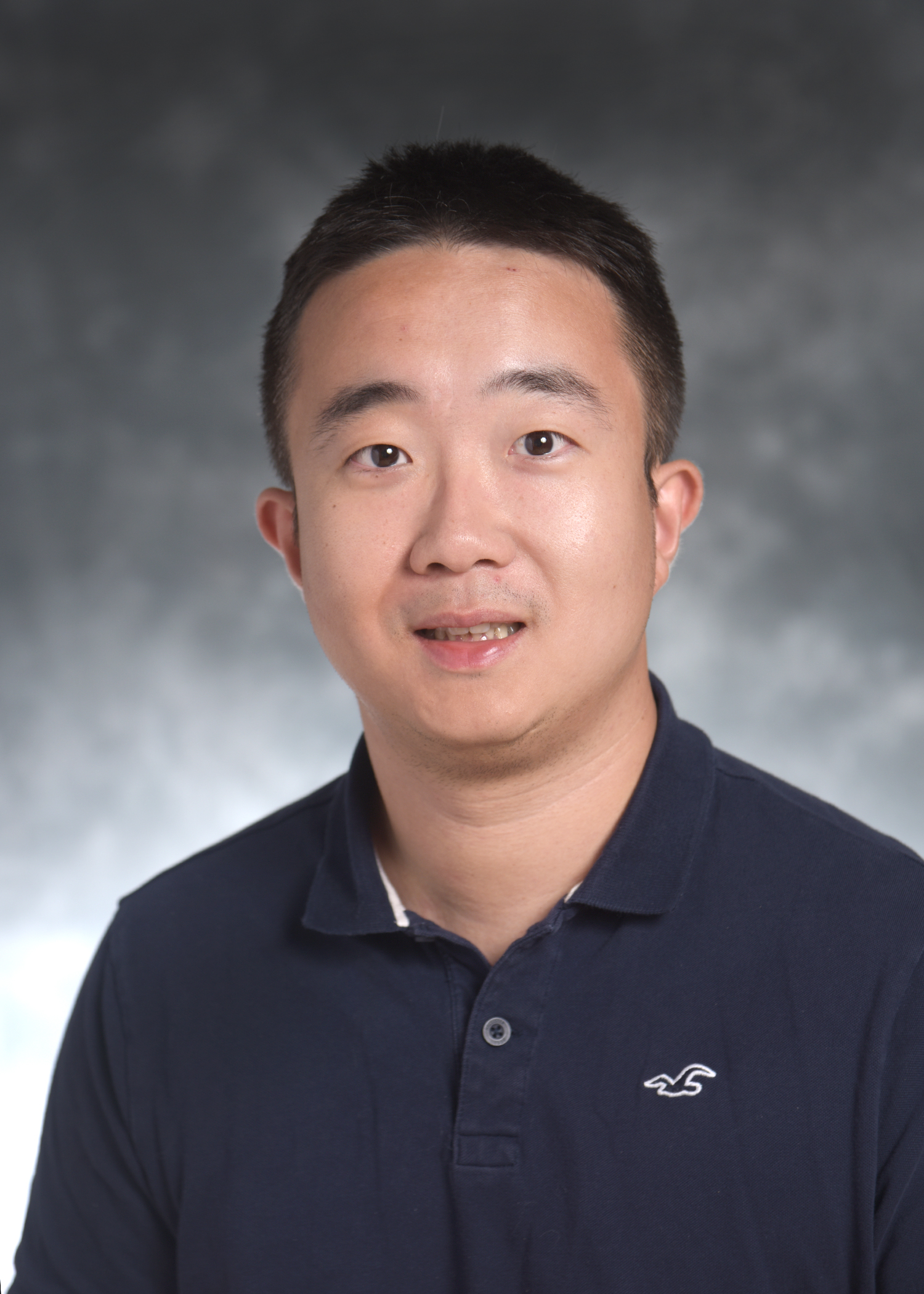}}]{Zihang Wei}received his M.Sc. degree in Civil Engineering from Texas A\&M University, College Station, TX, USA, in 2021. He is currently a Ph.D. candidate at the Zachry Department of Civil \& Environmental Engineering, Texas A\&M University. He is also working as a graduate research assistant at Texas A\&M Transportation Institute (TTI). His research interests include physics-informed modeling of traffic dynamics, traffic network control, equity in transportation systems, and traffic safety data analysis.
\end{IEEEbiography}
\vspace{-1.5em}
\begin{IEEEbiography}
[{\includegraphics[width=1in,height=1.25in,clip,keepaspectratio]{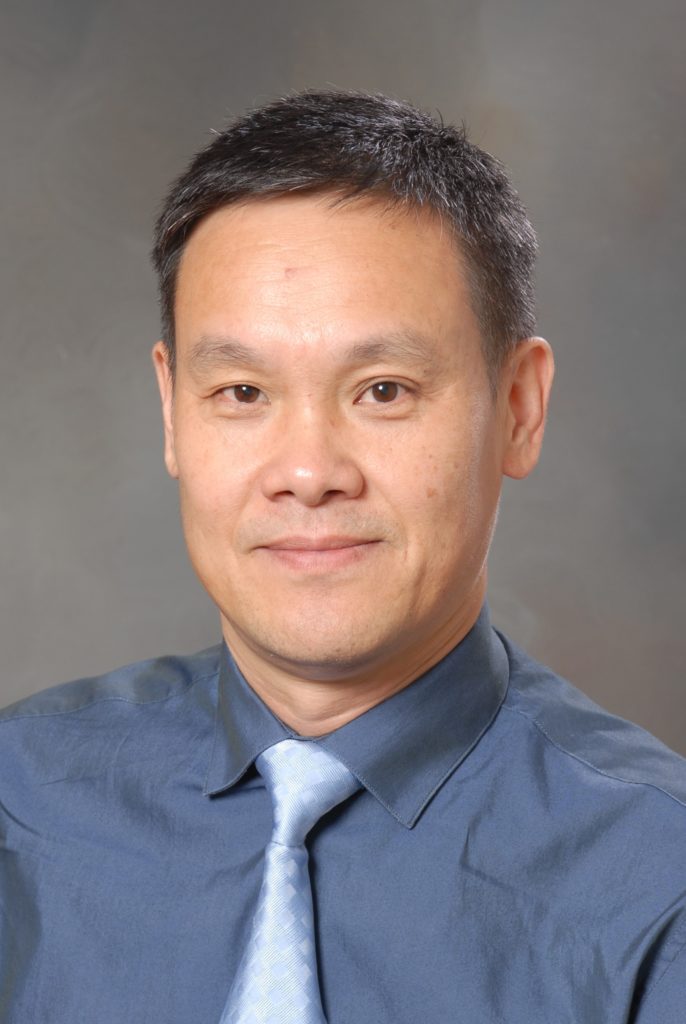}}]{Yunlong Zhang}received the B.S. degree in civil engineering and the M.Sc. degree in highway and traffic engineering from the Southeast University of China, Nanjing, China, in 1984 and 1987, respectively, and the Ph.D. degree in transportation engineering from the Virginia Polytechnic Institute, State University, Blacksburg, VA, USA, in 1996. He is currently a professor with the Zachry Department of Civil \& Engineering, Texas A\&M University, College Station, TX, USA. His research interests include transportation modeling and simulation, traffic operations, and Artificial Intelligence and data analytics applications in transportation.
\end{IEEEbiography}
\vspace{-1.5em}
\begin{IEEEbiography}
[{\includegraphics[width=1in,height=1.25in,clip,keepaspectratio]{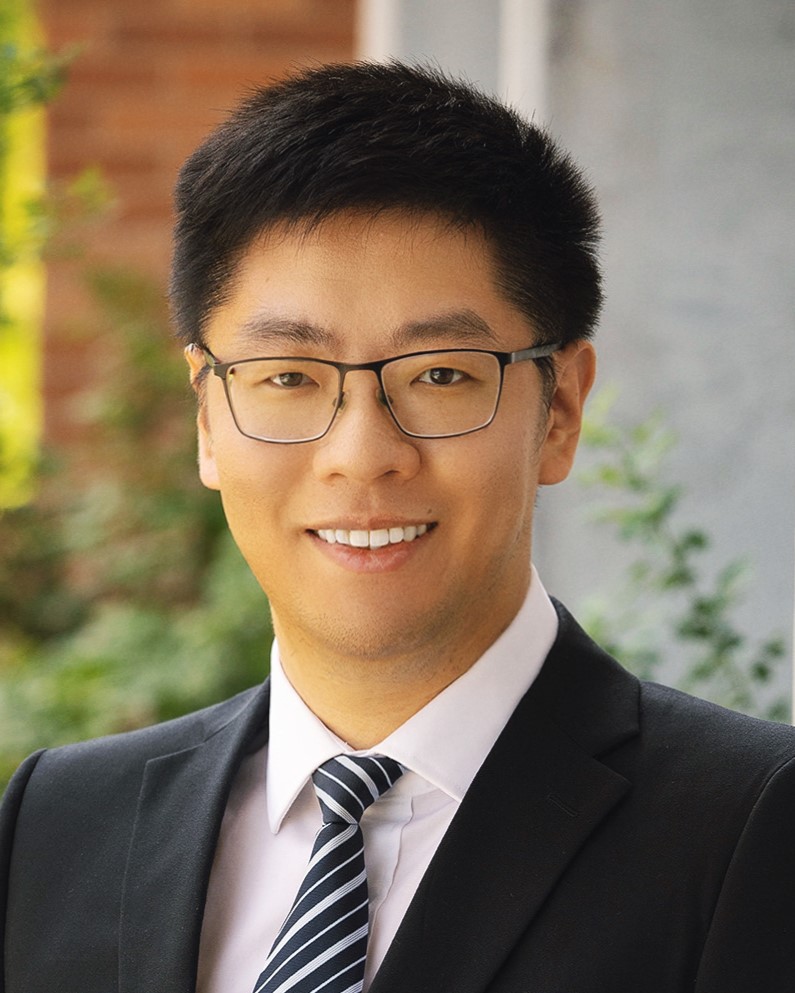}}]{Chenxi Liu}received the B.S. degree from Tsinghua University, Beijing, China, in 2017, and the M.S. and Ph.D. degrees from the University of Washington, Seattle, WA, USA, in 2020 and 2024, respectively. He is currently an Assistant Professor with the Department of Civil and Environmental Engineering, The University of Utah. He is an active member of ASCE, serving on both the Artificial Intelligence in Transportation Committee and the Active Transportation Committee. He also contributes to TRB standing Committee AED50 focsed on edge computing.His research interests include situation-aware customized machine intelligence to establish a connected and autonomous transportation system for safety and resiliency
\end{IEEEbiography}
\vspace{-1.5em}
\begin{IEEEbiography}
[{\includegraphics[width=1in,height=1.25in,clip,keepaspectratio]{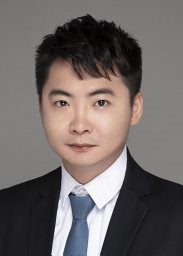}}]{Yang Zhou} received the Ph.D. degree in Civil and Environmental Engineering from University of Wisconsin Madison, WI, USA, in 2019, and the M.S. degree in Civil and Environmental Engineering from University of Illinois at Urbana-Champaign, Champaign, IL, USA, in 2015. He is currently an Assistant Professor in the Zachry Department of Civil and Environmental Engineering at Texas A\&M University. Before joining Texas A\&MUniversity, he was a postdoctoral researcher in Civil Engineering, University of Wisconsin Madison, WI, USA. He is currently a member in TRB traffic flow theory CAV subcommittee, network modeling CAV subcommittee, and American Society of Civil Engineering. His main research directions are connected automated vehicle robust control, interconnected system stability analysis, traffic big data analysis, and microscopic traffic flow modeling.
\end{IEEEbiography}
\vfill
\end{document}